\documentclass[journal,twocolumn,web]{article}
\usepackage[margin=0.8in]{geometry}
\usepackage{amsmath,amssymb,amsfonts}
\usepackage{algorithmic}
\usepackage{graphicx}
\usepackage{textcomp}
\usepackage{hyperref}

\usepackage{subfig}
\usepackage[detect-all, per-mode=symbol]{siunitx} % [detect-all] allows bold si units

\usepackage[style=ieee,url=false]{biblatex}
\date{\url{https://andersvra.github.io/research/coherenceaberration/}}

\begin{document}
\title{Coherence Based Sound Speed Aberration Correction — with clinical validation in fetal ultrasound}
\author{Anders Emil Vrålstad\thanks{Corresponding author, Anders Emil Vrålstad: anders.e.vralstad@ntnu.no.} \thanks{Department of Circulation and Medical Imaging, Norwegian University of Science and Technology (NTNU), 7491 Trondheim, Norway.}, Peter Fosodeder\thanks{GE Healthcare Womens Health, 4871 Tiefenbach, Austria}, Karin Ulrike Deibele\footnotemark[2] \thanks{Department of Women's Health, St. Olavs Hospital Trondheim University Hospital, Trondheim, Norway.}, Siri Ann Nyrnes\footnotemark[2] \thanks{Children's Clinic, St. Olav's Hospital, Trondheim University Hospital, Norway.}, \\ Ole Marius Hoel Rindal\thanks{Department of Informatics, University of Oslo, 0315 Oslo, Norway.}, Vibeke Skoura-Torvik\footnotemark[4] \thanks{Department of Clinical and Molecular Medicine, NTNU, 7491 Trondheim, Norway.\\This work was supported in part by the Center for Innovative Ultrasound Solutions (CIUS) and the Research Council of Norway under Project 237887.\\Project website with supplementary GIF images: \url{https://andersvra.github.io/research/coherenceaberration/}.}, \hspace{4pt}Martin Mienkina\footnotemark[3], and Svein-Erik Måsøy\footnotemark[2]}
\onecolumn
\maketitle
\begin{abstract}
The purpose of this work is to demonstrate a robust and clinically validated method for correcting sound speed aberrations in medical ultrasound. We propose a correction method that calculates focusing delays directly from the observed two-way distributed average sound speed. The method beamforms multiple coherence images and selects the sound speed that maximizes the coherence for each image pixel. 
The main contribution of this work is the direct estimation of aberration, without the ill-posed inversion of a local sound speed map, and the proposed processing of coherence images which adapts to \textit{in vivo} situations where low coherent regions and off-axis scattering represents a challenge. 
The method is validated \textit{in vitro} and \textit{in silico} showing high correlation with ground truth speed of sound maps. Further, the method is clinically validated by being applied to channel data recorded from 172 obstetric Bmode images, and 12 case examples are presented and discussed in detail. The data is recorded with a GE HealthCare Voluson Expert 22 system with an eM6c matrix array probe. The images are evaluated by three expert clinicians, and the results show that the corrected images are preferred or gave equivalent quality to no correction (\SI{1540}{\meter\per\second}) for 72.5\% of the 172 images. In addition, a sharpness metric from digital photography is used to quantify image quality improvement. The increase in sharpness and the change in average sound speed are shown to be linearly correlated with a Pearson Correlation Coefficient of 0.67.
\end{abstract}
\twocolumn
\section{Introduction}
\label{sec:Introduction}
The speed of sound is a fundamental parameter in medical ultrasound imaging. It determines how the ultrasound system focuses, and deviations from the true value will impact the image quality of the system. Knowledge of the true sound speed of the imaging medium can be used as a biomarker with diagnostic value, and it can be used to correctly focus and remove aberrations from sound speed errors \cite{ali_aberration_2023}. A conventional ultrasound system assumes a sound speed of \SI{1540}{m/s} and this may deviate from the true value, for example in patients with a higher percentage of fatty tissue, as fat is known to have a sound speed of \SI{1450}{\meter\per\second} \cite{azhari_appendix_2010}. The speed of sound parameter on the scanner has been shown to impact the duration and completion of clinical examinations \cite{dashe_maternal_2009,chauveau_improving_2018}. For fetal ultrasound diagnostics, improving the contrast and sharpness is in particular important. Parent counseling and their choices depends on accurate information regarding prognosis based on the ultrasound examination. Furthermore, early detection and diagnosis, for example of fetal congenital heart defects, have an impact on neonatal survival and morbidity \cite{yeo_fetal_2018}. In this paper, we included a group of mothers with high body mass index (BMI) to challenge the technology for sound speed aberration correction \cite{sujan_randomised_2023}.

Extensive research has been conducted to perform sound speed reconstruction \cite{stahli_bayesian_2021,beuret_windowed_2024} and aberration correction \cite{lambert_ultrasound_2022,ali_distributed_2022} in pulse echo ultrasound, as demonstrated in a recent review article by Ali et al. \cite{ali_aberration_2023}. A group of sound speed estimators optimizes an image quality metric by beamforming using multiple sound speed values \cite{napolitano_sound_2006, ali_local_2022, ali_distributed_2022}. This procedure can be performed by employing synthetic transmit and receive focusing with the same received signals, and thereby correct aberrations both on receive and transmit retrospectively. Some methods are based on first estimating the local sound speed map and, secondly, computing the travel time to use to focus the received signals \cite{ali_local_2022,ali_distributed_2022}. These types of estimators measure the average sound speed between the transmit aperture, imaging points, and the receive aperture. Solving for the local sound speed map is ill-posed because there exist multiple local sound speed maps that produce the same average sound speed. 

In this article, we propose a sound speed aberration correction method that optimizes a coherence factor and computes the focusing delays directly from the average sound speed map. Thus, we avoid the ill-posed problem of estimating the local sound speed map. The method is motivated by the fact that the calculation of the average sound speed map is similar to the travel-time calculation when assuming wave propagation in straight rays. The method assumes that refraction can be neglected due to small deviations ($\pm$10\%) in tissue sound speed. 

The method includes a scan-grid compensation for an unbiased estimate of the sound speed \cite{vralstad_sound_2024}. Negative sidelobes caused by adaptive beamforming and large dynamic variations of backscatter reflectivity produce a challenge when maximizing the coherence. This is mitigated by thresholding the coherence and using rank filtering to spatially expand the main lobe. We use focused transmit and harmonic imaging to suppress reverberations and transmit sidelobes to improve the accuracy of our sound speed estimate \cite{ali_local_2022}. \textit{In vivo} channel data signals are stored for offline beamforming using a 2D matrix array. Matrix arrays are known to be better suited for aberration estimation and correction than 1D apertures \cite{trahey_evaluation_1991, liu_comparison_1995}. REFoCUS beamforming is used because synthetic aperture focusing is independent of the assumed sound speed and focus position during transmission \cite{bottenus_recovery_2018}. Our sound speed estimate is therefore unbiased against the sound speed assumed on transmission and allows for two-way aberration correction using focused transmission \cite{bottenus_recovery_2018}.

\section{Theory}
\label{sec:theory}
\begin{figure}
    \centering
    \includegraphics[trim ={0cm 0cm 0cm 0cm}, clip, width=.5\linewidth]{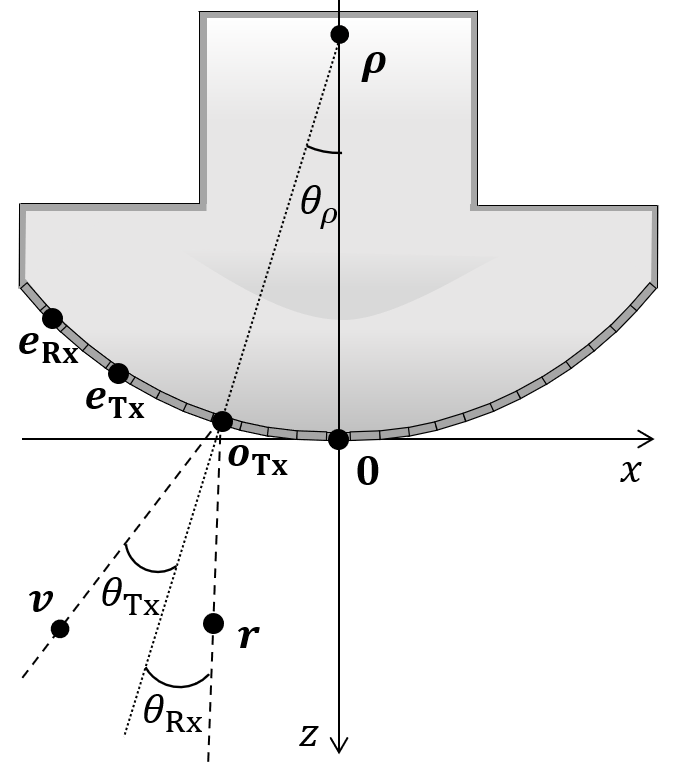}
    \caption{Visualization of imaging setup with the coordinate system origin $\mathbf{0} = \left[0,0,0\right]^T$. The figure shows the transmit origin $\mathbf{o_{\text{Tx}}}$, the receive element $\mathbf{e_{\text{Rx}}}$, the transmit element $\mathbf{e_{\text{Tx}}}$, the transmit focus position (virtual source) $\mathbf{v}$ and pixel position $\mathbf{r}$.}
    \label{fig:coordinates}
\end{figure}

The coordinate system and an example of the positioning of the imaging setup are visualized in Fig.~\ref{fig:coordinates}. Point positions are notated as 3D Euclidean vectors with bold vector notation (\textit{e.g.} $\mathbf{r} = \left[r_x,r_y,r_z\right]^T$). 

\subsection{Wave propagation model}
\label{sec:wave_prop_model_theory}
\begin{figure}[!ht]
    \centering
    \subfloat[\label{fig:aberration_ambiguity_NO} No aberration.]{
        \includegraphics[trim ={0.5cm 2cm 30.5cm 0cm}, clip, width=.3\linewidth]{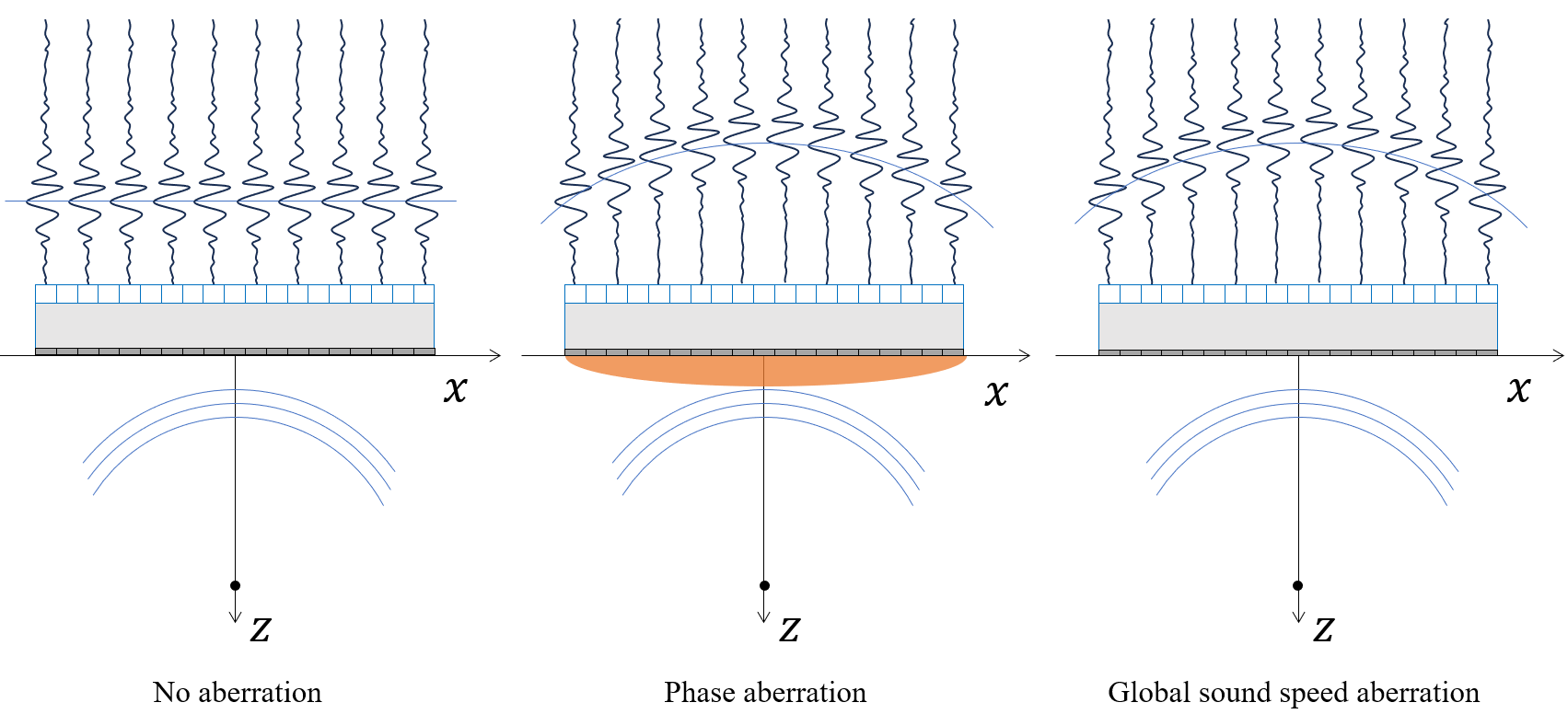}}
    \subfloat[\label{fig:aberration_ambiguity_PHASE} Lens aberration.]{
        \includegraphics[trim ={15.5cm 2cm 15.5cm 0cm}, clip, width=.3\linewidth]{Figures/aberration_ambiguity.png}}
    \subfloat[\label{fig:aberration_ambiguity_GLOBAL} Wrong global sound speed aberration.]{
        \includegraphics[trim ={31cm 2cm 0cm 0cm}, clip, width=.3\linewidth]{Figures/aberration_ambiguity.png}}
    \caption{Illustration of received signal from a point scatter after focusing. The received signals are similar for (b) and (c) even though (b) is aberrated by a lens and (c) is aberrated by an incorrect global sound speed.}
    \label{fig:aberration_ambiguity}
\end{figure}

The model chosen to estimate and correct sound speed errors assumes that the wave propagation from the active transmit aperture to a pixel and back to the receivers acts as a global sound speed with individual values for individual pixels. The sound speed values in the map are the average of the different straight ray paths between the $N$ active transmit elements, the pixels, and the $M$ receive elements. This spatially distributed average sound speed $c_{\text{avg}}(\mathbf{r})$ is expressed as
\begin{equation}
    c_{\text{avg}}(\mathbf{r}) =\frac{1}{MN}\sum_{m=1}^{M} \sum_{n=1}^{N} c_{\text{har}}(\mathbf{r},\mathbf{e}_{\text{Tx}}(n))+c_{\text{har}}(\mathbf{r},\mathbf{e}_{\text{Rx}}(m)),
    \label{eq:2way_average_speed}
\end{equation}
where $c_{\text{har}}(\mathbf{r},\mathbf{e})$ is the one-way harmonic average of the local sound speed from element position $\mathbf{e}$ to pixel position $\mathbf{r}$, given by
\begin{equation}
    c_{\text{har}}(\mathbf{r},\mathbf{e})^{-1} =\frac{1}{\|\mathbf{r}-\mathbf{e}\|_2} \displaystyle\int\limits_{0}^{\|\mathbf{r}-\mathbf{e}\|_2}\frac{dl}{c_{\text{local}}(\mathbf{e}+l(\mathbf{r}-\mathbf{e})/\,\|\mathbf{r}-\mathbf{e}\|_2)},
    \label{eq:harmonic_1way_average}
\end{equation}
where $\|\cdot\|_2$ is the Euclidean norm and $l$ is the straight path length integrated over. 
The assumption neglects refraction and will interpret any hyperbolic delay aberration as an error in sound speed. This can lead to an aberration ambiguity as illustrated in Fig.~\ref{fig:aberration_ambiguity}. This simplified model of wave propagation captures a first-order approximation of the propagation in a medium with heterogeneous sound speed. Refraction and phase aberrations are here treated as second-order effects. Accounting for refraction is more complicated as it is sensitive to the accuracy of the estimated local sound speed map \cite{brevett_comparison_2023}.
The aberration correction is performed directly by beamforming with the average sound speed map $c_{\text{avg}}$.

Note that the spatial variation of the average sound speed maps is expected to be smooth for longer path lengths due to the cumulative integration in \eqref{eq:harmonic_1way_average}. 

\subsection{Beamforming with average sound speed map}
\label{sec:beamforming_theory}
During beamforming we assume the simplified wave propagation model from Section~\ref{sec:wave_prop_model_theory} and use a spatial weighted REFoCUS \cite{vralstad_universal_2024} implementation given as 
\begin{equation}
    b(\mathbf{r}) = \sum_{m=1}^{M} \sum_{k=1}^{K} \sum_{n=1}^{N} w(\mathbf{r},m,k,n)\, s(k,m,t(\mathbf{r},m,k,n)),
    \label{eq:generalized_refocus_beamformer}
\end{equation}
where $s(k,m,t)$ is the analytic channel data signal, and $M$, $K$ and $N$ are the number of active receive elements, transmit events and transmit elements, respectively. The two-way TOF $t$ is expressed as
\begin{equation}
        \begin{split}
            t(\mathbf{r},m,k,n)&=\frac{\|\mathbf{v}(k)-\mathbf{o_{Tx}}(k)\| - \|\mathbf{v}(k)-\mathbf{e_{Tx}}(n,k)\|}{c_0}\\
            &+\frac{\|\mathbf{r}-\mathbf{e_{Tx}}(n,k)\|}{c_{\text{avg}}(\mathbf{r})
            }+\frac{\|\mathbf{r}-\mathbf{e_{Rx}}(m)\|}{c_{\text{avg}}(\mathbf{r})},
    \end{split}
    \label{eq:refocus_two_way_TOF}
\end{equation}
where the first term is the reversed applied transmit delay and uses the assumed sound speed $c_0$. The second and third terms are the transmit and receive TOF respectively, and use the average sound speed which is a function of pixel position $\mathbf{r}$.

\subsubsection{Pixel-grid compensation}
The spatial mapping from time to space in a pixel-based beamformer depends on the speed of sound. This becomes important for the comparison of the focusing quality from pixel to pixel because it can bias the estimation if not taken into account \cite{vralstad_sound_2024,ahmed_spatial_2024}. The bias is mitigated by compensating the scan grid to maintain the geometry assumed on transmission (computed with $c_0$). The geometry compensation of the pixel depth is a simple scaling by the sound speed ratio starting at the receive beam origin, given as
\begin{equation}
    \|\mathbf{r}_{}-\mathbf{o_{Tx}}\| = \frac{c_{\text{avg}}(\mathbf{r})}{c_{0}}\|\mathbf{r}_{0}-\mathbf{o_{Tx}}\|.
    \label{eq:pixel_depth_correction}
\end{equation}
The angle of pixels changes with a change of sound speed according to Snell-Descartes law of refraction as
\begin{equation}
    \theta_\text{Rx}(\mathbf{r}) = \arcsin{\left(\frac{c_{\text{avg}}(\mathbf{r})}{c_{0}}\sin{\left(\theta_{\text{Rx0}}(\mathbf{r})\right)}\right)},
    \label{eq:pixel_steering_correction}
\end{equation}
where $\theta_{\text{Rx0}}(\mathbf{r})$ is the angle in the coordinate system computed with the sound speed $c_0$ and $\theta_{\text{Rx}}(\mathbf{r})$ computed with $c_{\text{avg}}(\mathbf{r})$.
The compensation makes all structures stationary while varying $c_{\text{avg}}(\mathbf{r})$. This is also beneficial because geometric measurements and standardized clinical indices are established with the scan grid obtained with $c_0 = 1540$\unit{\meter\per\second}. 

\subsection{Average sound speed estimation}
\label{sec:focusing_quality_theory}
The average sound speed can be found by retrospectively focusing the same data set using multiple global sound speeds and choosing a sound speed value for every pixel that optimizes a focusing quality metric \cite{ali_local_2022, ali_distributed_2022,ahmed_spatial_2024}. We employ a similar optimization using the coherence factor $C_R$ in \eqref{eq:rigby_coherence}, suggested in \cite{rigby_method_1999}.
\begin{equation}
    C_R(\mathbf{r}) = \frac{\left|\sum_{m=1}^{M}\hat{b}(\mathbf{r},m)\right|}{\sum_{m=1}^{M} \left|\hat{b}(\mathbf{r},m)\right|},
    \label{eq:rigby_coherence}
\end{equation}
where  $\hat{b}(\mathbf{r},m)$ is the beamsum, that is, the delay-and-summed channel data, summed over $K$ transmit events and $N$ transmit elements, given as
\begin{equation}
     \hat{b}(\mathbf{r},m) =\sum_{k=1}^{K}\sum_{n=1}^{N} w(\mathbf{r},m,k,n)\,s(k,m,t(\mathbf{r},m,k,n)).
     \label{eq:presum}
\end{equation}
The numerator in \eqref{eq:rigby_coherence} is the coherent sum of the received signals $s$ and the denominator is the incoherent sum.
\subsection{Optimal sound speed selection near a strong scatterer}
\label{sec:coherent_scatter}
Adaptive coherence-based beamformers are known to effectively suppress side lobes in regions near strong scatterers  \cite{rindal_dark_2017}. Coherence maximization in these regions may choose an incorrect sound speed because incorrect sound speeds will suppress sidelobes less efficiently.
%hasegawa_phase_2014

\begin{figure}
    \centering
    \subfloat[\label{fig:no_percentile} No percentile filter.]{
        \includegraphics[trim ={0cm 0cm 19.5cm 2cm}, clip, width=0.5\linewidth]{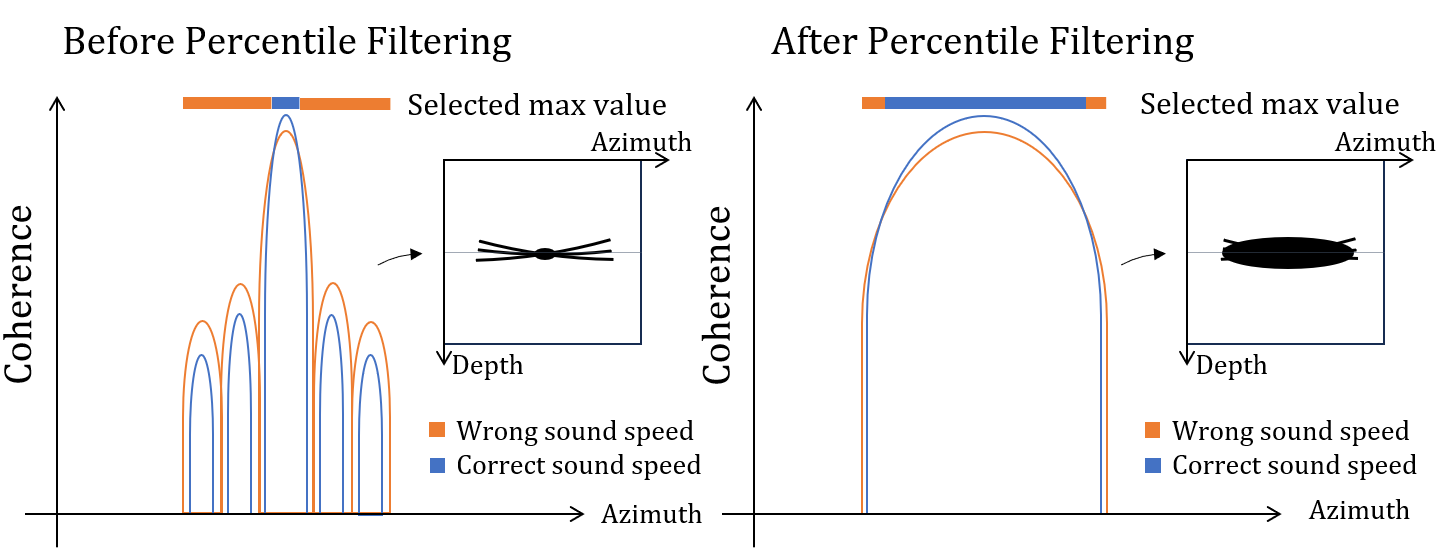}}
    \subfloat[\label{fig:with_percentile} With percentile filter.]{
        \includegraphics[trim ={19.5cm 0cm 0cm 2cm}, clip, width=0.5\linewidth]{Figures/percentile.png}}
    \caption{Illustration of Point Spread Function (PSF) beamformed with correct and wrong speed of sound. The bar on top in (a) shows that the max value is chosen to be the PSF using correct sound speed in the mainlobe, but incorrect sound speed in the sidelobes. After percentile filtering in (b) the correct sound speed is chosen in the sidelobes as well.}
    \label{fig:percentile_filter}
\end{figure}

Maximization of coherence in a region adjacent to a strong scatterer is then susceptible to an erroneous speed of sound selection because this will actually lead to a higher sidelobe level here (see illustration in Fig.~\ref{fig:percentile_filter}).

Such problems can be mitigated by spatial expanding of the coherence values in each of the coherence images before selecting the maximum across different sound speeds. Spatial expansion can be implemented using an image processing technique called percentile or rank filtering \cite{hodgson_properties_1985}. This filtering operation is performed with a 2D kernel sliding over the image picking a high percentile coherence value within the kernel. The 50th percentile and 100th percentile versions of such a filter are equivalent to a median filter and image dilation, respectively. The 90th percentile is used for the methods in this paper. Figure \ref{fig:percentile_filter} illustrates how the percentile filter works by practically expanding the main lobe of the Point Spread Function (PSF) of the system, thus ensuring that the maximum coherence value is also selected off-axis from a strong scatterer. 
\section{Method}
\label{sec:Method}
The sound speed correction is implemented in the python beamforming library vbeam \cite{kvalevag_vbeam_2023}. The distributed sound speed aberration correction is performed in five steps:
\begin{enumerate}
    \item Calculate multiple coherence factor images with constant sound speed (1440, 1450,...,\SI{1640}{\meter\per\second}).
    \item Threshold, smooth and rank filter the coherence images.
    \item Select the sound speed with highest coherence. 
    \item Median filter and smooth the sounds speed map.
    \item Beamform the channel data using the average sound speed map from 4).
\end{enumerate}
The steps are described in more detail below, and the parameter values and kernel sizes for all filters are provided in Tables~\ref{tab:kwave_correction_settings} and ~\ref{tab:eM6C_settings}.

The coherence images in Step 1) are beamformed from harmonic filtered channel data using the expression in \eqref{eq:rigby_coherence} and use the TOF calculation in \eqref{eq:refocus_two_way_TOF} with constant sound speed values $c(\mathbf{r})=c$. The pixel-grid compensation in \eqref{eq:pixel_depth_correction}-\eqref{eq:pixel_steering_correction} is applied for the different coherence images. 

In Step 2), we first replace some of the coherence values inside a mask. The mask is the logical union of three pixel regions; pixels where the coherence values are less than 0.2, a depth region in the extreme nearfield and at the scan sector borders (given as a percentage of the scan sector width in Tables~\ref{tab:kwave_correction_settings} and \ref{tab:eM6C_settings}). The coherence values inside the mask are replaced by the mean value of the coherence values outside the mask; $C_R[P]\leftarrow\overline{C_R[\neg P]}$, where $P$ is the logical mask, $\neg$ is the negation operator, and $\overline{\text{overline}}$, indicate the mean value. 
Secondly, the coherence images are smoothed with a 2D Gaussian window to average measurement noise. A moving average is also applied for each pixel over the sound speed dimension, smoothing the coherence as a function of the sound speed. Finally, a 2D rank filtering, described in Section~\ref{sec:coherent_scatter}, mitigates the influence from off-axis scatterers. 

In Step 3) the sound speed with maximum coherence, after the processing in Step 2), is selected for each pixel independently. 
In Step 4) a 2D median filter is applied to remove outliers from the map. Secondly, the average sound speed map is smoothed with a 2D kernel size that increases linearly with depth. This increase is done because the average sound speed map is expected to be smoother at higher depths. 

The aberration correction in Step 5) is performed by beamforming the channel data using the expression in \eqref{eq:generalized_refocus_beamformer} and the TOF calculated using the average sound speed map from Step 3). 

Final displayed images were processed and visualized with software running the same image processing libraries as those used on the Voluson Expert 22 scanner. In this way, the change in image quality from the proposed sound speed correction method can be evaluated on the basis of clinically relevant images.

\subsection{Validation in controlled settings}
The validity of the distributed aberration correction method is tested \textit{in silico} and \textit{in vitro}.
\subsubsection{\textit{In-silico}}
Three K-wave simulations \cite{treeby_k-wave_2010} using a linear scan and focused transmissions are executed in a medium with different local sound speed maps. The simulation assumes \SI{1540}{\meter\per\second} during transmit focusing and images a speckle scene with point targets. The channel data from the simulations are recorded and average sound speed is estimated with the proposed method. To evaluate the precision of the estimate, the ground-truth average sound speed is calculated from the true local sound speed maps using \eqref{eq:2way_average_speed}. The assumed sound speed during transmit is \SI{1540}{\meter\per\second} and 21 sound speed values (1440, 1450, ... \SI{1640}{\meter\per\second}) are used in the estimation process. The simulation settings and processing settings are found in Table~\ref{tab:kwave_settings} and  Table~\ref{tab:kwave_correction_settings} respectively.
\begin{table}[h!]
    \centering
    \caption{K-wave simulation setup.}
    \begin{tabular}{c|c|c}
         Parameter &                Value &                 Units \\
         \hline
         Number of elements &       128 &                  - \\
         Array pitch &              400 &                    $\mu$m \\
         \hline
         Pulse &                    Gaussian 2-cycles &     - \\
         Center frequency &         4 &                     MHz \\
         \hline
         Transmit focus depth &     30&                     mm \\
         \hline
         Grid spacing &             100 &                   $\mu$m \\
         Grid dimension &           1080$\times$1024&      pixels \\
         Medium density &           1020  &                 kg/m$^3$          
    \end{tabular}
    \label{tab:kwave_settings}
\end{table}
\begin{table}[h]
    \centering
    \caption{Aberration correction settings for the K-wave simulation.}
    \begin{tabular}{c|c|c}
         Parameter &                Value &                 Units \\
         \hline
         \textbf{Step 2)} & & \\
         Extreme nearfield depth &  3 &                    mm \\
         Valid percentage of sector & 100 &                  \% \\ 
         2D Gaussian kernel size &  6$\times$5         &   mm$\times$mm\\
         Moving average length &    3 &                     -\\
         Percentile kernel size &   8$\times$2         &   mm$\times$mm\\
         \hline 
         \textbf{Step 4)} & & \\
         Median filter size &       4$\times$6         &   mm$\times$mm\\
         2D kernel size &           18 $\times$8       &  mm$\times$mm\\
         Lateral increase rate  &      0.3            &  mm/mm \\
         Axial increase rate  &      0.3 &                  mm/mm \\
    \end{tabular}
    \label{tab:kwave_correction_settings}
\end{table}
\subsubsection{\textit{In vitro}}
Two sets of channel data are collected from the same CIRS general-purpose phantom, assuming different sound speeds during transmission. The phantom has a sound speed of \SI{1540}{\meter\per\second}. Data were collected using the Voluson Expert 22 scanner and the eM6C curvilinear matrix-array probe from GE HealthCare (GE HealthCare Women's Health Ultrasound, Zipf, Austria). The Expert 22 system allows for storage of channel data for offline processing. The eM6c use Sub-Aperture technology to limit the number of channels \cite{haugen_ultrasound_2005}. The two recordings are collected to test the sound speed estimators on the sound speed assumption during transmission. The probe position is stationary for the two recordings. The settings used in the aberration correction process for the \textit{in vitro} and \textit{in vivo} images are given in Tab.~\ref{tab:eM6C_settings}.
\begin{table}[h]
    \centering
     \caption{Aberration correction settings for the eM6C probe.}
    \begin{tabular}{c|c|c}
         Parameter &                Value &                 Units \\
         \hline
         \textbf{Step 2)} & & \\
         Extreme nearfield depth &  10 &                    mm \\
         Valid percentage of sector & 80 &                  \% \\ 
         2D Gaussian kernel size &  10$\times$5         &   degrees$\times$mm\\
         Moving average length &    3 &                     -\\
         Percentile kernel size &   5$\times$3         &   degrees$\times$mm\\
         \hline 
         \textbf{Step 4)} & & \\
         Median filter size &       10$\times$10         &   degrees$\times$mm\\
         2D kernel size &           20 $\times$10       &   degrees$\times$mm\\
         Lateral increase rate  &      0.803            &  degrees/mm \\
         Axial increase rate  &      0.6 &                  mm/mm \\
    \end{tabular}
    \label{tab:eM6C_settings}
\end{table}
\subsection{Clinical validation of the method}
Between May and December 2023, twelve pregnant women were included and 13 fetal ultrasound examinations were performed. Two of the women had normal prepregnancy BMI, while six were overweight and four were obese (BMI ranged from 19.2-\SI{36.1}{\kilogram\per\meter^2}). A total of 172 Bmode images of clinical standard views for fetal biometry and echocardiography were obtained in pregnancy weeks 19 and/or 32-33. The images were captured for the following measurements; BiParietal Diameter (BPD), Mean Abdominal Diameter (MAD), Femur Length (FL), Four chamber view (4CV), Pulmonary Valve Diameter (PVD) and Aortic Valve Diameter (AVD). Data were collected with the eM6C probe on the Voluson Expert 22 system at St. Olavs Hospital, Trondheim University Hospital, Norway. Participants in the before the beginning trial were invited and provided written informed consent \cite{sujan_randomised_2023}. The project was approved by the Regional Committee for Medical and Health Research Ethics, REC central. The data were captured using a focused walking aperture transmit sequence assuming \SI{1540}{\meter\per\second} on transmit. The images are aberration corrected with the proposed method and the settings used in this process are similar to the \textit{in vitro} experiment given in Tab.~\ref{tab:eM6C_settings}.
\subsubsection{Clinical evaluation}
\label{sec:clinical_study}
One midwife with 8 years of experience and two medical doctors with 15 and 25 years of experience, all from the Fetal Medicine Department at St. Olavs hospital, evaluated the clinical relevance of the sound speed correction. The clinicians were presented with two images, one using \SI{1540}{\meter\per\second} and one corrected for aberration, and they selected their preferred image. The clinicians also had the option to mark the image quality as similar. The image pairs were randomly shuffled in the evaluation.

\subsubsection{Tenengrad sharpness}
\label{sec:tenengrad}
Tenengrad sharpness is a robust and accurate measure for focusing quality and performs well in digital photography \cite{groen_comparison_1985, mir_extensive_2014}. The Tenengrad measure is the magnitude of the Sobel filtered image. Because the correct speed of sound mostly affects the lateral focusing quality in pulse echo ultrasound, we use the lateral Sobel filter expressed as
\begin{equation}
    \mathbf{G}(\mathbf{r}) =  \begin{bmatrix}
                        1 & 0 & -1\\
                        2 & 0 & -2\\
                        1 & 0 & -1
                    \end{bmatrix} * \mathbf{A}(\mathbf{r}),
    \label{eq:Sobel_filtering}
\end{equation}
where the image $\mathbf{A}$ is convolved with the matrix to produce the image $\mathbf{G}$. The Tenengrad measure $F$ is calculated by summing the magnitude of the filtered image $\mathbf{G}$ as
\begin{equation}
    F = \sum_{\mathbf{r}\in\text{ROI}} |\mathbf{G}(\mathbf{r})|
    \label{eq:Global_Tenengrad}
\end{equation} where the sum of pixels $\mathbf{r}$ is within a Region Of Interest (ROI) consisting of all angles within the transmitted beam sector and a depth axis beginning at \SI{10}{mm} to avoid measuring contributions from near-field artifacts. The percentage increase of image sharpness between the image corrected for sound speed and the uncorrected image is given as
\begin{equation}
    \kappa = 100\left(\frac{F_\text{corrected}}{F_\text{uncorrected}}-1\right).
    \label{eq:percentage_increase}
\end{equation}
We expect the sharpness to be correlated with the proposed sound speed correction method. To test this, we calculate the Mean Absolute Deviation (MAD) of the estimated average sound speed map from \SI{1540}{\meter\per\second} and measure the Pearson Correlation Coefficient (PCC) between the MAD and the Tenengrad image quality increase $\kappa$.

\section{Results}
\label{sec:Results}
\subsection{Simulated experiments}
\begin{figure}[!ht]
    \centering
    \subfloat[\label{fig:silico_global_1480_Bmode} Correctly \\focused Bmode of \\\SI{1480}{\meter\per\second} simulation.]{
        \includegraphics[trim ={2.5cm .5cm 2.5cm 1.5cm}, clip, width=.3\linewidth]{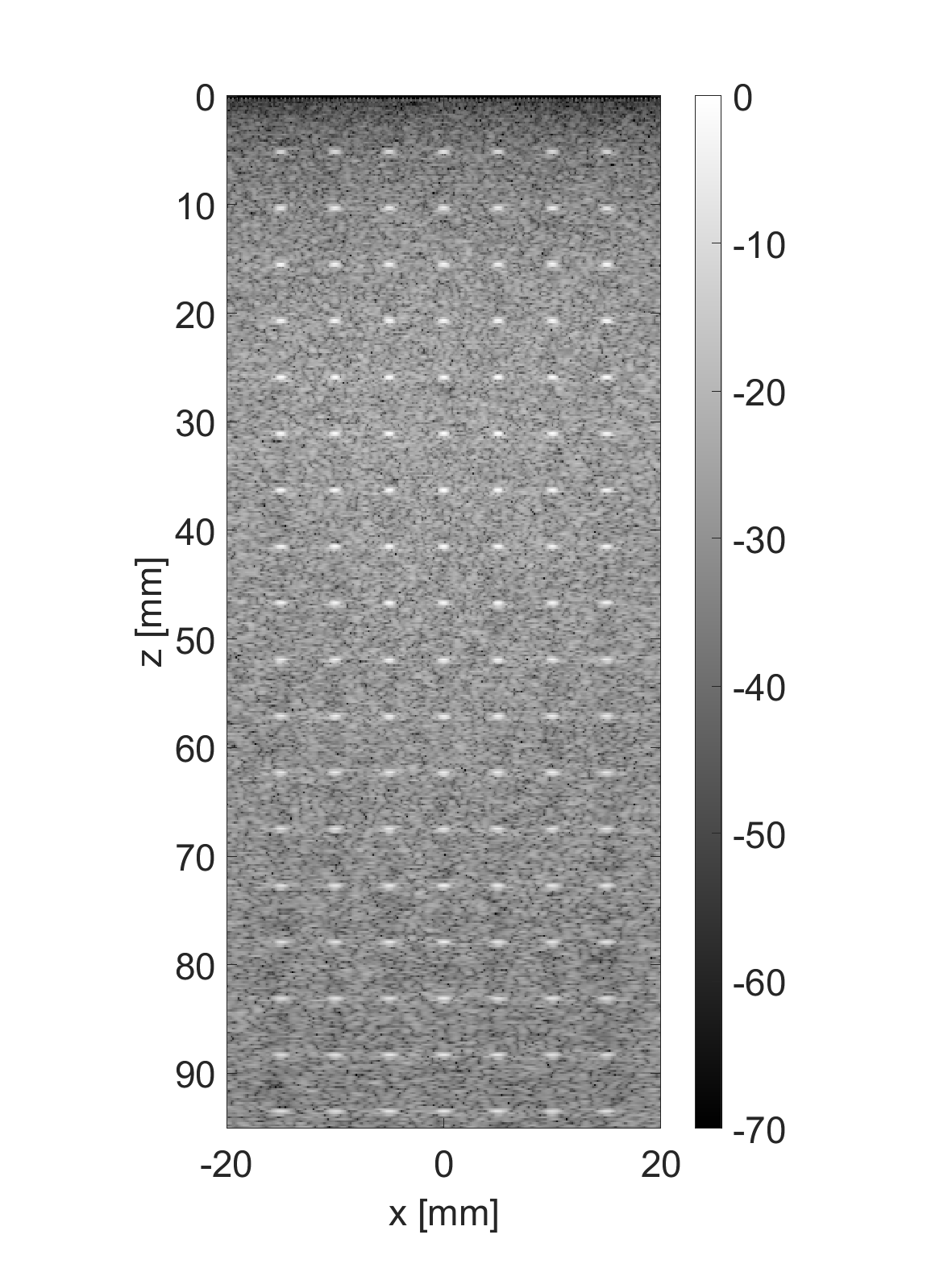}}
    \subfloat[\label{fig:silico_global_1480} Estimated map \\of constant \SI{1480}{\meter\per\second} \\map. MAE=\SI{4.78}{\meter\per\second}.]{
        \includegraphics[trim ={2.5cm .5cm 2.5cm 1.5cm}, clip, width=.3\linewidth]{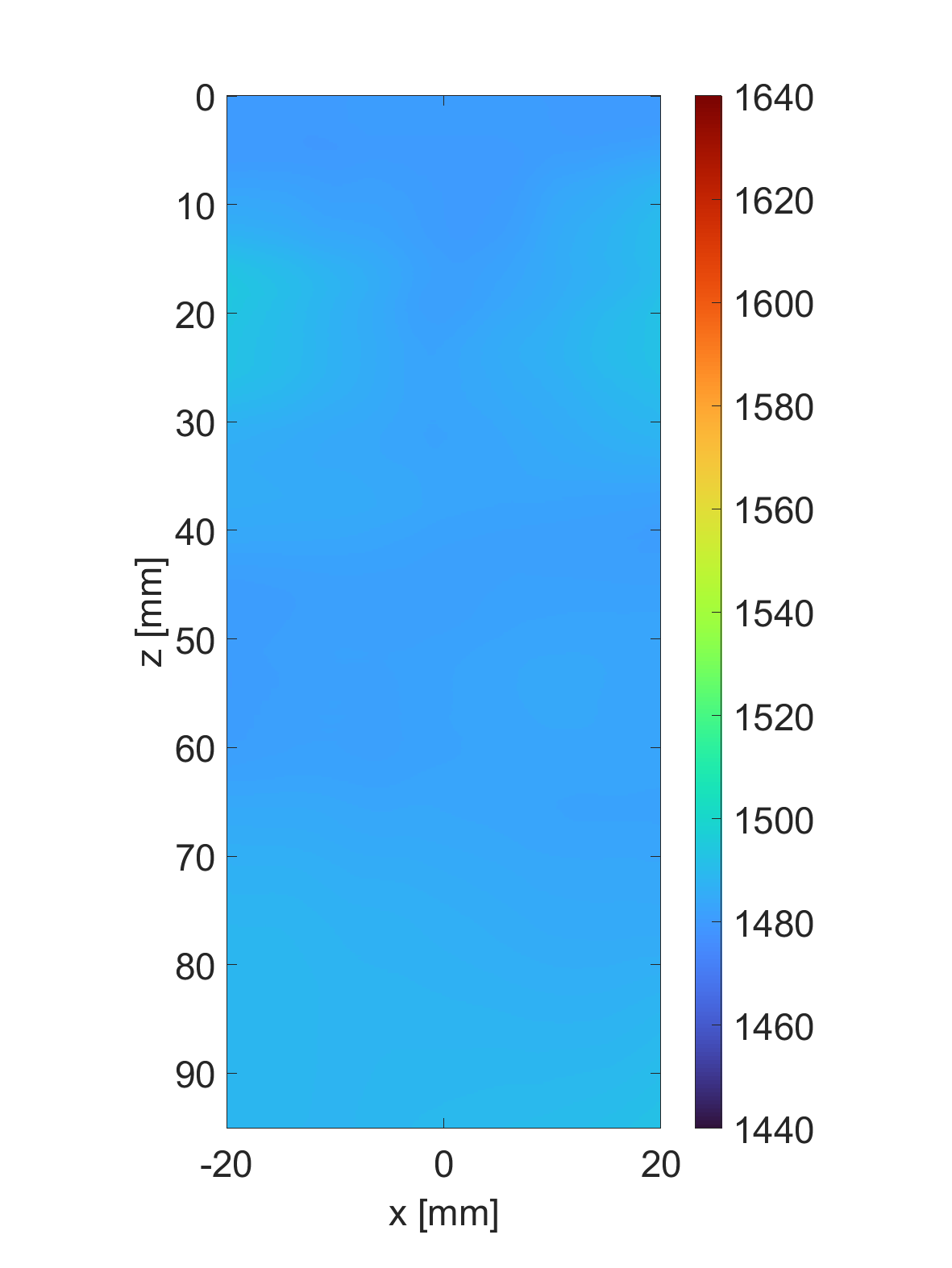}}
    \subfloat[\label{fig:silico_global_1610}  Estimated map \\of constant \SI{1610}{\meter\per\second} \\map. MAE=\SI{4.74}{\meter\per\second}.]{
        \includegraphics[trim ={2.5cm .5cm 2.5cm 1.5cm}, clip, width=.3\linewidth]{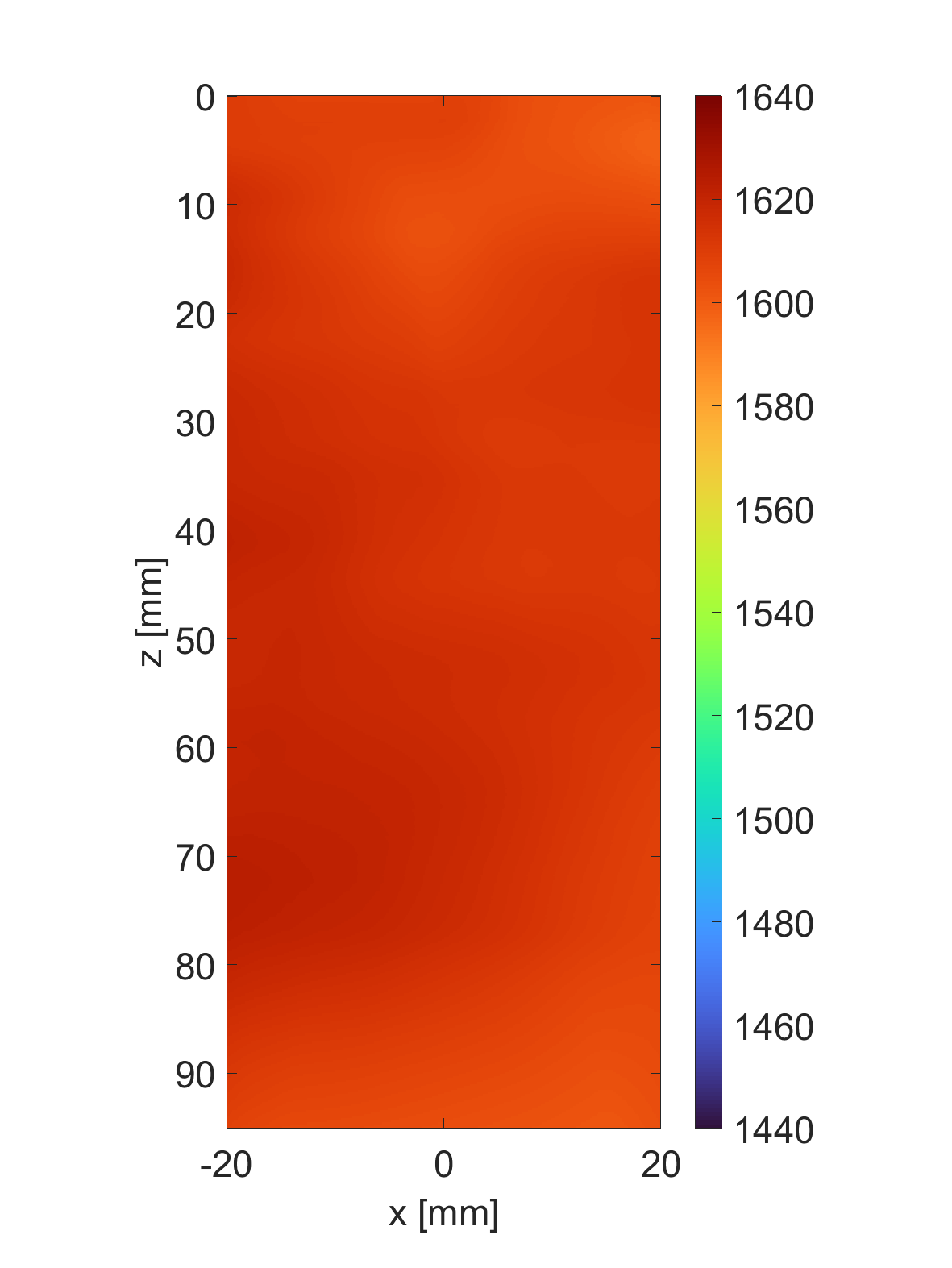}}
    \caption{Estimated average sound speed map from \textit{in silico} speckle phantoms with point targets and constant global sound speeds. The Mean Absolute Error (MAE) from the ground truth is written in the subcaptions.}
    \label{fig:global_silico}
\end{figure}

\begin{figure}[!ht]
    \centering
    \subfloat[\label{fig:silico_layered_local} Local map.]{
        \includegraphics[trim ={2.5cm .5cm 2.5cm 1.5cm}, clip, width=.3\linewidth]{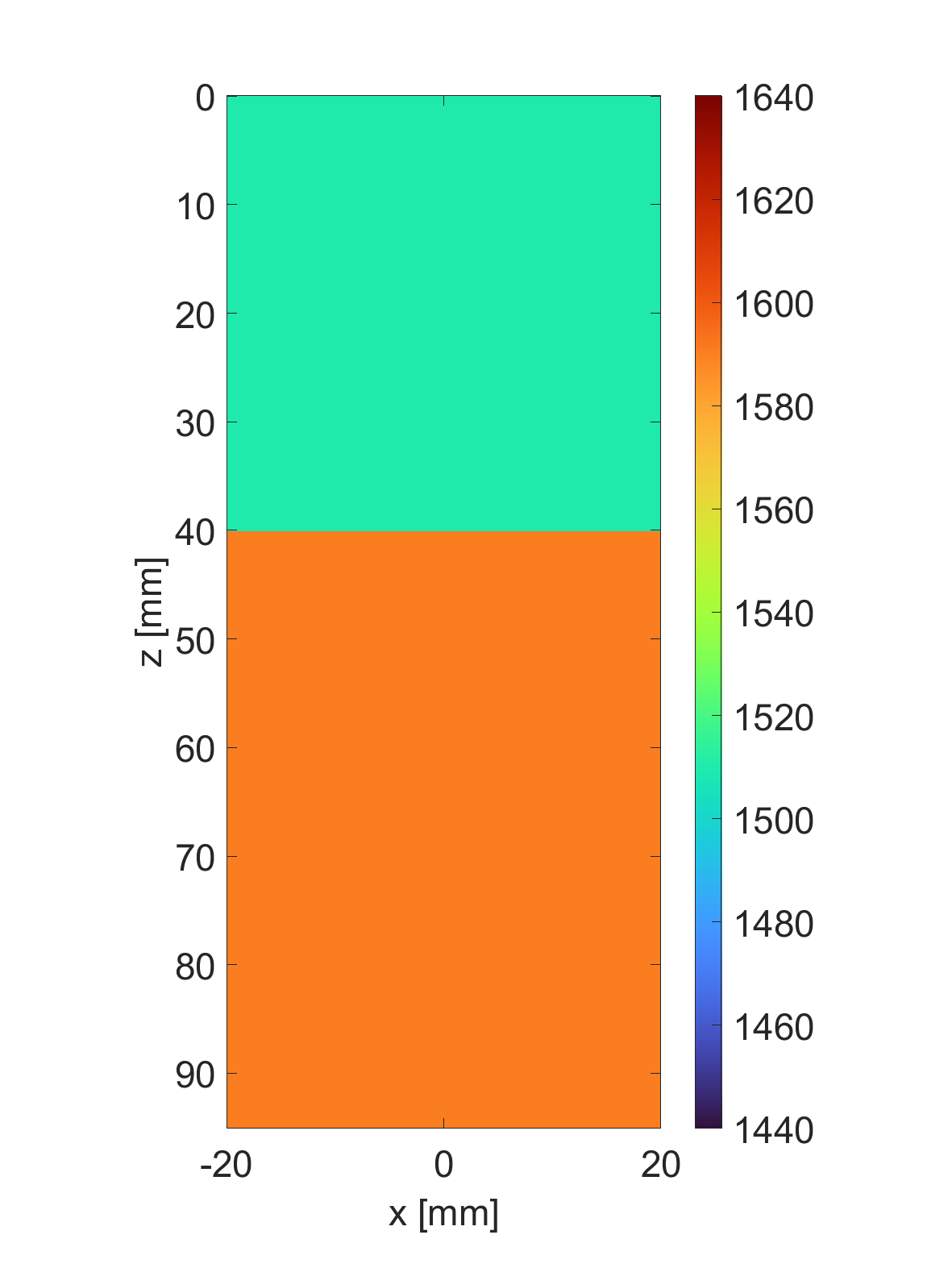}}
    \subfloat[\label{fig:silico_layered_true_avg} True average.]{
        \includegraphics[trim ={2.5cm .5cm 2.5cm 1.5cm}, clip, width=.3\linewidth]{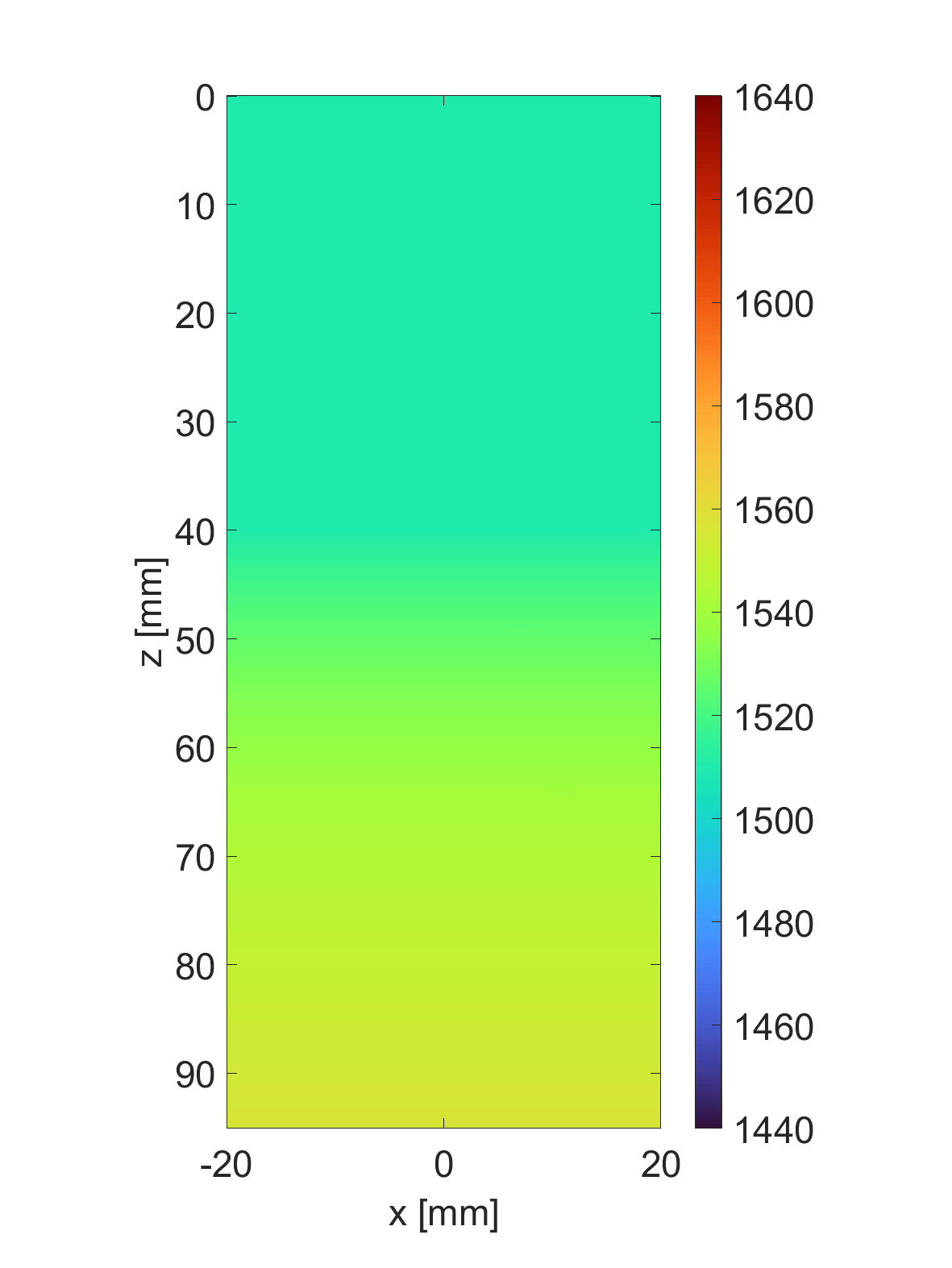}}
    \subfloat[\label{fig:silico_layered_est_avg} Estimated average. MAE=4.31\unit{\meter\per\second}.]{
        \includegraphics[trim ={2.5cm .5cm 2.5cm 1.5cm}, clip, width=.3\linewidth]{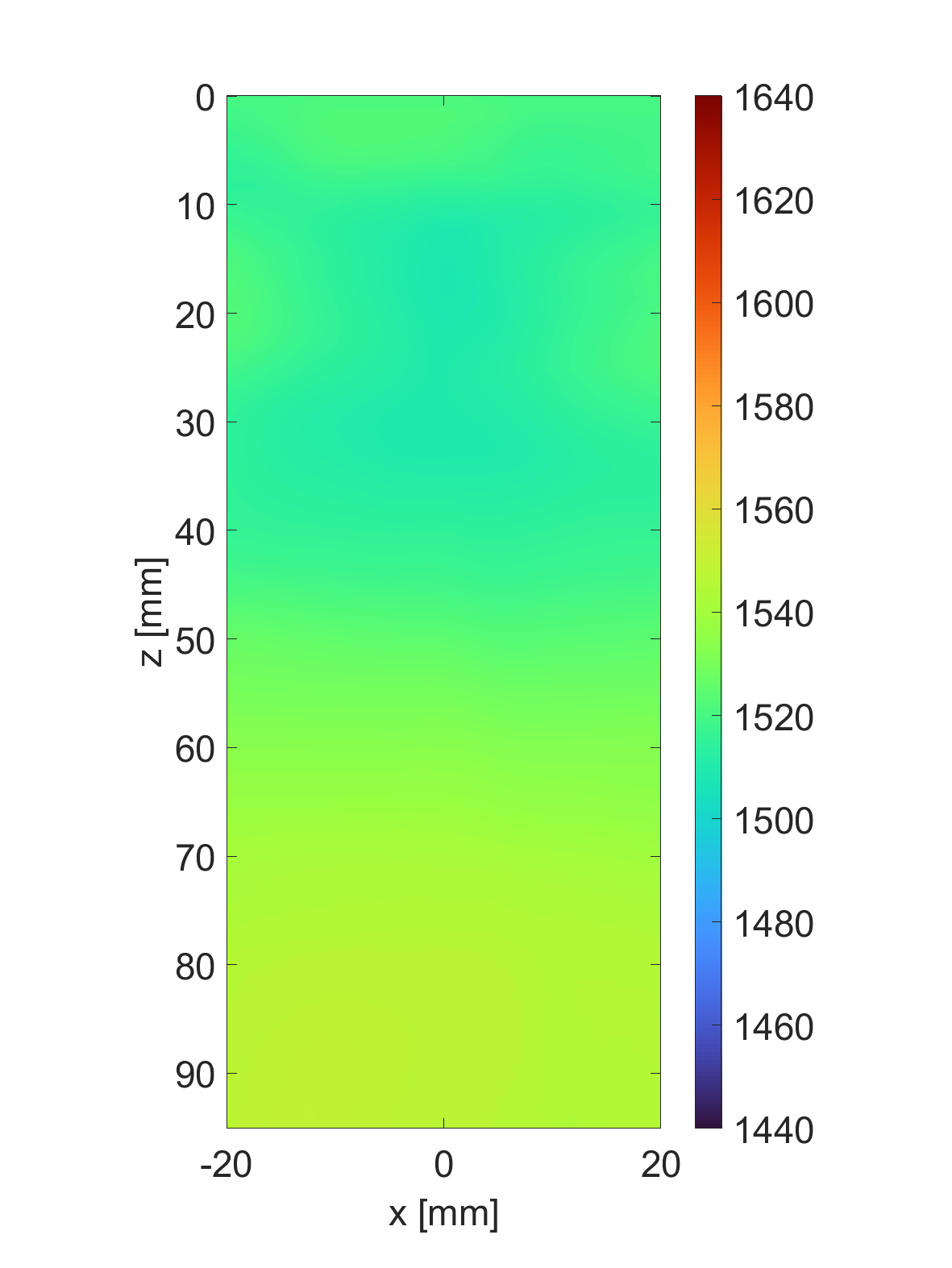}}
    \caption{Local, true average and estimated average sound speed map from \textit{in silico} phantom with speckle, point targets  and a layered sound speed map. The Mean Absolute Error (MAE) from the ground truth is found in the subcaption.}
    \label{fig:layered_silico}
\end{figure}

Speckle, point targets and different sound speed maps are simulated in K-wave. Fig.~\ref{fig:silico_global_1480_Bmode} shows the Bmode of the medium, beamformed using the true sound speed. Fig.~\ref{fig:global_silico} shows the estimated sound speed maps for true constant \SI{1480}{\meter\per\second} and \SI{1610}{\meter\per\second} values. Fig.~\ref{fig:layered_silico} shows the true local, true averaged and the estimated sound speed maps for a layered medium. The true average map in Fig.~\ref{fig:silico_layered_true_avg} is calculated from the local map in Fig.~\ref{fig:silico_layered_local} using the wave propagation model assumed in \eqref{eq:2way_average_speed}. The estimated map follows the procedures in Section~\ref{sec:Method}.


\subsection{\textit{In vitro} experiments}
\begin{figure}[!ht]
    \centering
    \subfloat[\label{fig:phantom_map_1540} Estimated sound speed map. \\assuming \SI{1540}{\meter\per\second} on transmit. \\MAE = \SI{1.05}{\meter\per\second}.]{
        \includegraphics[trim ={1cm 3.5cm 0cm 4cm}, clip, width=.5\linewidth]{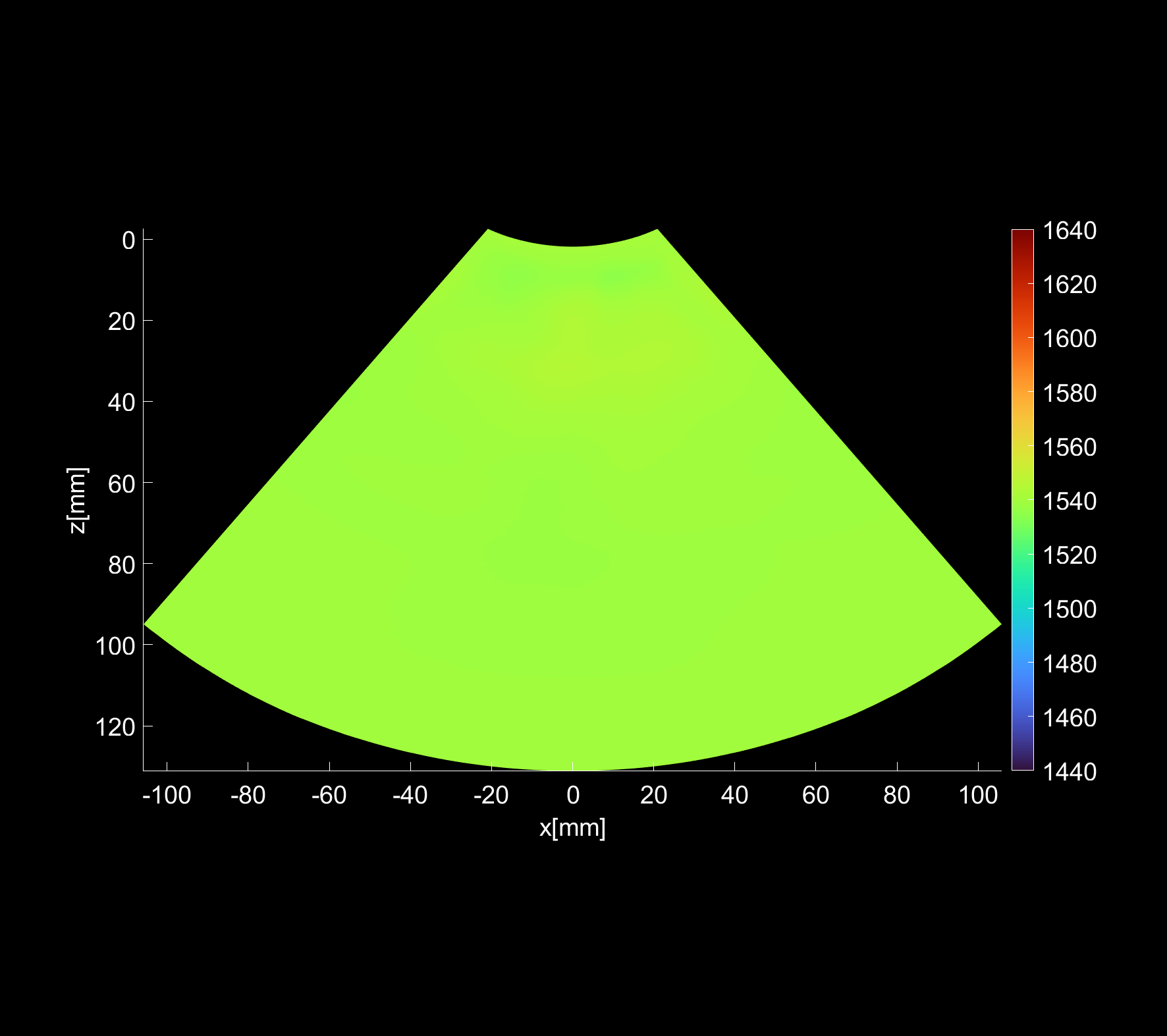}}
    \subfloat[\label{fig:phantom_bmode_1540} Aberration corrected Bmode \\assuming \SI{1540}{\meter\per\second} on transmit.]{
        \includegraphics[trim ={1cm 3.5cm 0cm 4cm}, clip, width=.5\linewidth]{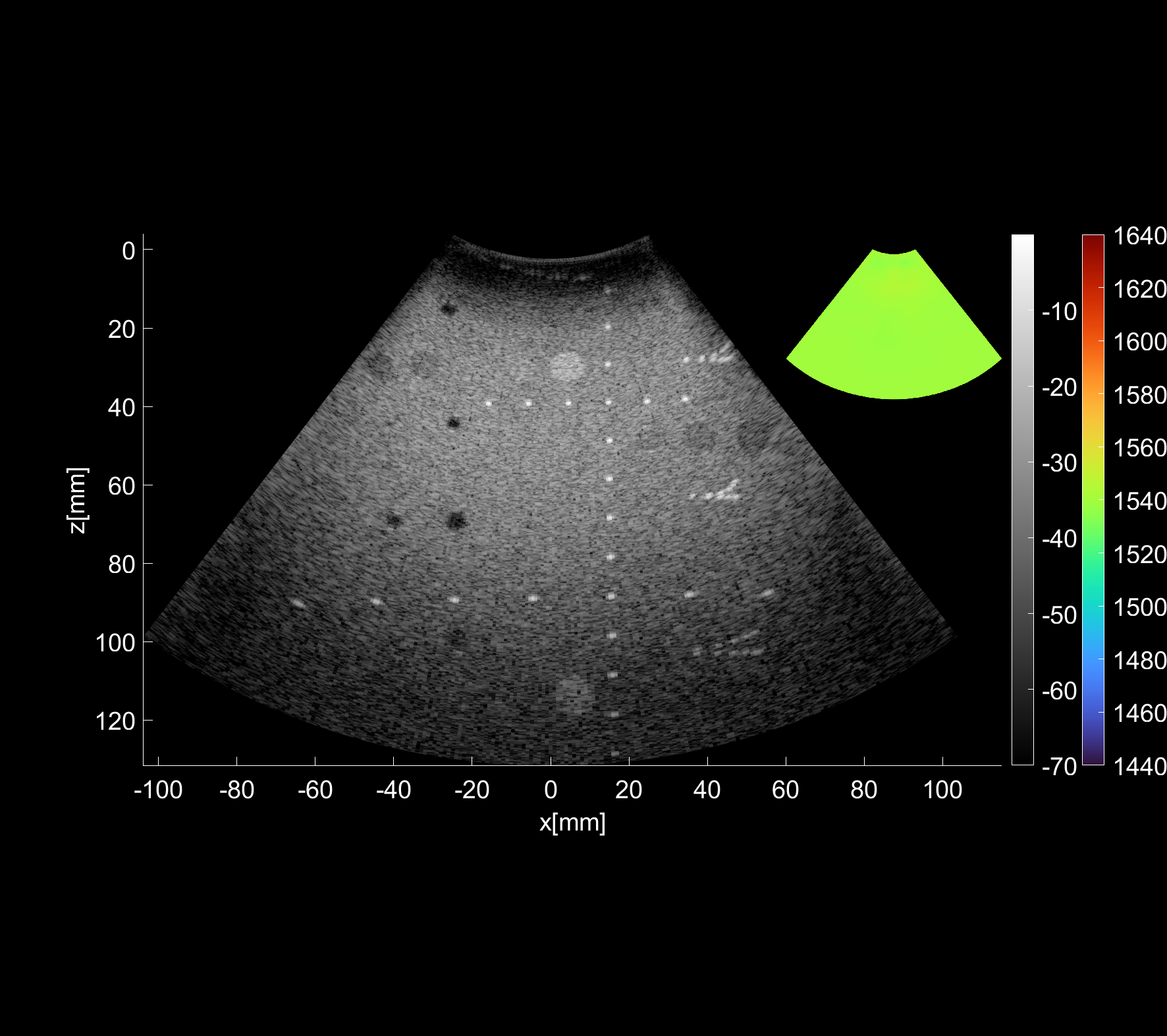}}

    \subfloat[\label{fig:phantom_map_1480} Sound speed map estimated \\assuming \SI{1480}{\meter\per\second} on transmit.  \\MAE = \SI{1.26}{\meter\per\second}.]{
        \includegraphics[trim ={1cm 3.5cm 0cm 4cm}, clip, width=.5\linewidth]{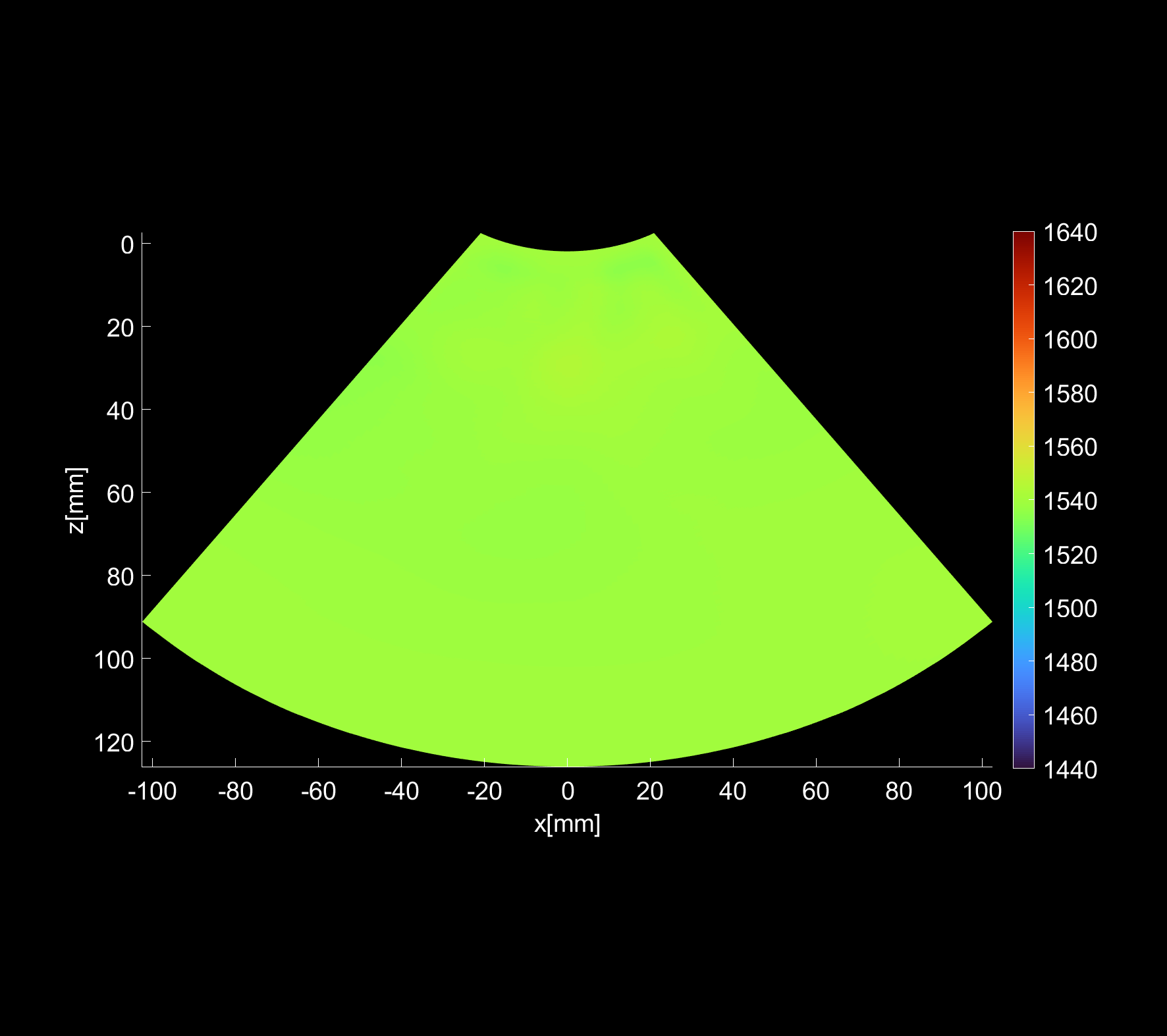}}
    \subfloat[\label{fig:phantom_bmode_1480} Aberration corrected Bmode \\assuming \SI{1480}{\meter\per\second} on transmit.]{
        \includegraphics[trim ={1cm 3.5cm 0cm 4cm}, clip, width=.5\linewidth]{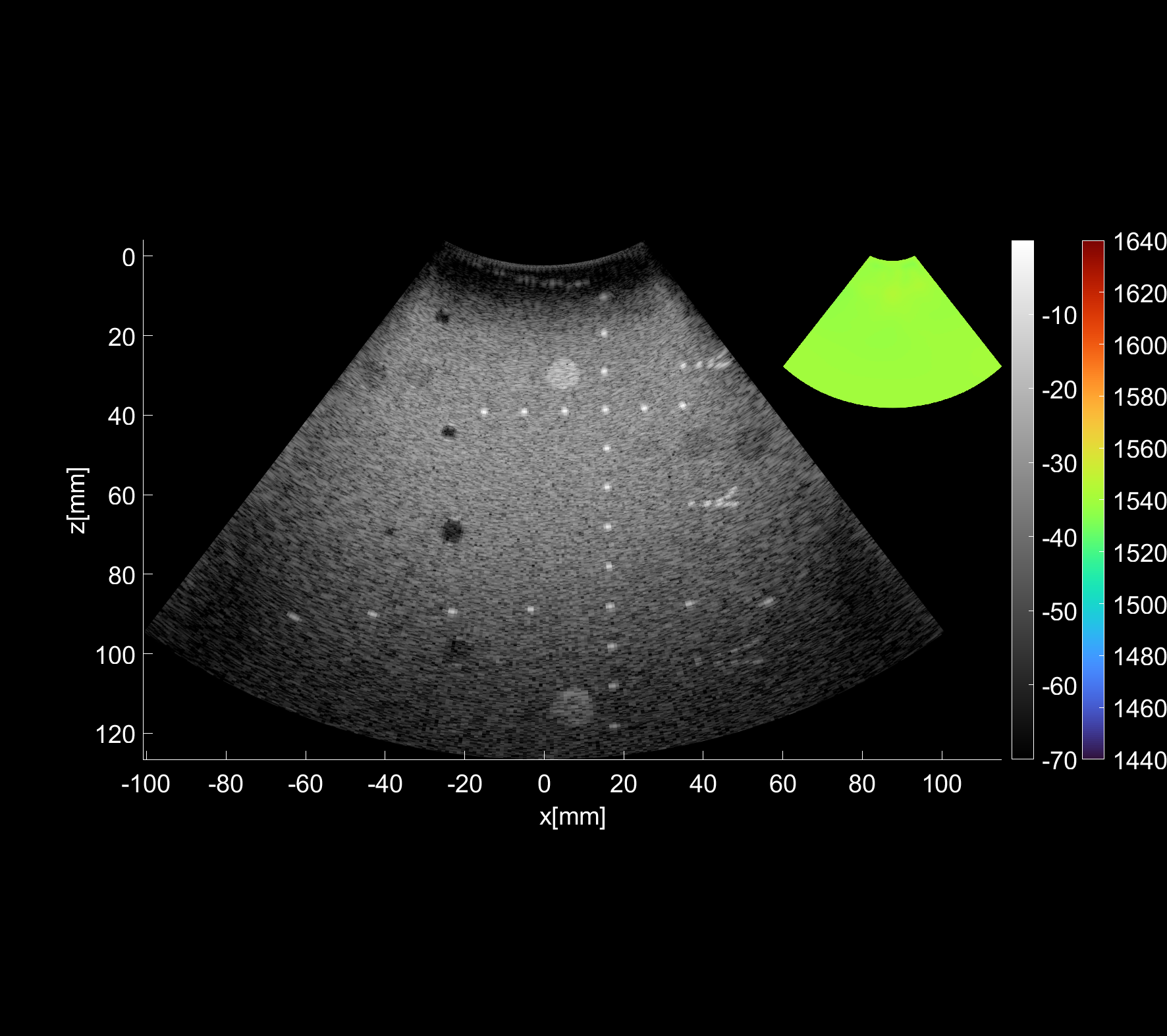}}
    \caption{Estimated average sound speed maps and Bmode images of two collections of a CIRS \SI{1540}{\meter\per\second} phantom. The recordings differ in the assumed sound speed during transmission.}
    \label{fig:invitro1}
\end{figure}

The Voluson Expert 22 scanner has the option to set the sound speed to different constant values (OTI parameter). Two channel data sets from the same CIRS phantom are collected assuming \SI{1540}{\meter\per\second} and \SI{1480}{\meter\per\second}. The effective focus position and the scan grid will change between the two acquisitions as a consequence of changing sound speed \cite{vralstad_sound_2024}. The estimated average sound speed map and Bmode images are shown in Fig.~\ref{fig:invitro1}.

\subsection{Clinical \textit{in vivo} validation}
All the 172 pairs of corrected and uncorrected clinical fetal Bmode images have been evaluated by three clinicians and image quality improvement has been estimated using the Tenengrad image quality metric $F$ from Section~\ref{sec:tenengrad}. 

\begin{figure}
    \centering
    \includegraphics[trim ={0cm 0cm 0cm 0cm}, clip, width=\linewidth]{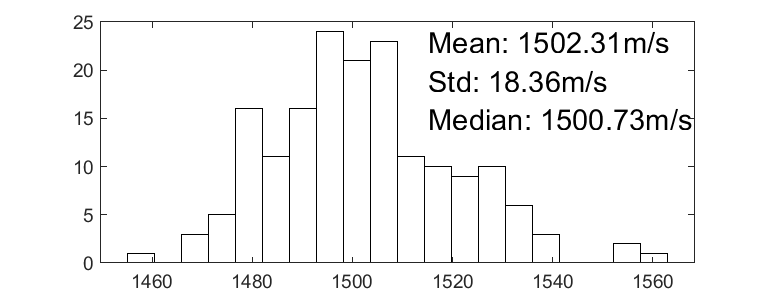}
    \caption{Distribution of estimated global speed of sound values in 172 \textit{in vivo} fetal images.}
    \label{fig:Single_sound_speed_histogram}
\end{figure}

\begin{figure}
    \centering
    \includegraphics[trim ={0cm 0cm 0cm 0cm}, clip, width=\linewidth]{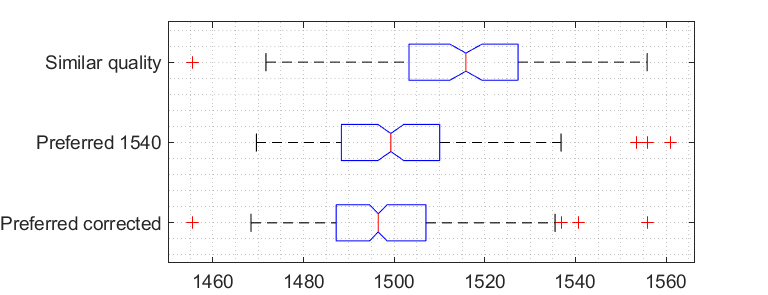}
    \caption{Distribution of the estimated global speed of sound values in the population of images label by the clinical evaluators. The notch indicate 95\% confidence of the median. The median of "similar quality" is statistically different from the others by the Wilcoxon rank sum test.}
    \label{fig:boxplot_sound_speeds}
\end{figure}

The average value of the estimated average sound speed maps provides global values for each image in the dataset. The distribution of the global values over the entire dataset is shown in Fig.~\ref{fig:Single_sound_speed_histogram}. 
Table~\ref{tab:clinical_study} presents the evaluations of each clinician and shows that the proposed method is preferred or creates a quality similar to the uncorrected image in 72.5\% (50.6\%+21.9\%) of the cases. Box plots of the global average sound speed estimates labeled by the evaluators are shown in Fig.~\ref{fig:boxplot_sound_speeds}. 
\begin{table*}[]
    \centering
    \caption{Clinical evaluation of image quality.}
    %\begin{scriptsize}
    \begin{tabular}{c|c|c|c}
            & Prefer corrected & Prefer \SI{1540}{\meter\per\second} & Similar quality \\
            \hline 
        Evaluator 1 &  95 & 47  & 30 \\
        Evaluator 2 &  66 & 29  & 77 \\
        Evaluator 3 &  100 & 66  & 6 \\
        \hline
        Sum & 261 & 142 & 113\\
        Percentage & 50.6\% & 27.5\% & 21.9\% \\   
    \end{tabular}
    %\end{scriptsize}
    \label{tab:clinical_study}
\end{table*}

\begin{figure}[!ht]
    \centering
    \includegraphics[trim ={0.3cm 0cm 0.1cm 0.3cm}, clip, width=\linewidth]{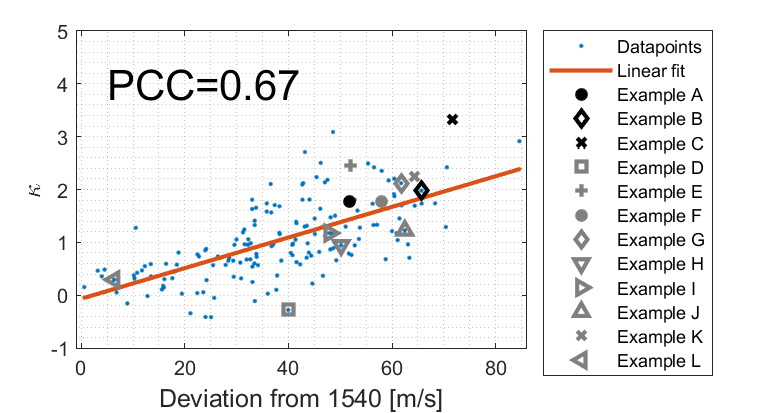}
    \caption{Percentage increase of image quality by correcting for sound speed instead of using \SI{1540}{\meter\per\second}. The quality is measured using the image power after lateral Sobel filtering. Examples A-C are visualized in Figs.~\ref{fig:invivo1}
    -\ref{fig:invivo3} and D-L (in gray) are found in supplementary materials. }
    \label{fig:quanititative_sharpness}
\end{figure}

The percentage increase in image sharpness, $\kappa$, is calculated according to \eqref{eq:percentage_increase}. The increase is plotted as a function of the estimated sound speed deviation in Fig.~\ref{fig:quanititative_sharpness}. The Pearson Correlation Coefficient (PCC) is measured to be 0.67. 

\begin{table*}[]
    \centering
    \caption{Tabular values and evaluation answers from 12 case examples.}
    \begin{footnotesize}
    \begin{tabular}{l|c|c|c|c|c|c|c|c|c|c|c|c}
        Example    & A&  B & C & D & E & F & G & H & I & J & K & L \\
            \hline 
        Global speed estimate [\si{\meter\per\second}] &  1488 & 1474 & 1468 & 1500 & 1488 & 1482 & 1478 & 1490 & 1492 & 1478 & 1475  & 1534\\ 
        Sharpness increase ($\kappa$) &  1.78 &  1.98 &  3.33 &  -0.28 &  2.46 &  1.77 &  2.12 &  0.95 &  1.17 &  1.23 &  2.25 & 0.30 \\
        Corrected preferred &  3 &  2 &  3 &  1 &  2 &  3 &  1 &  3 &  3 &  0 &  3 & 0\\
        Uncorrected preferred &  0 &  1 &  0 &  1 &  1 &  0 &  2 &  0 &  0 &  3 &  0  & 0\\
        Similar quality &  0 &  0 &  0 &  1 &  0 &  0 &  0 &  0 &  0 &  0 &  0 & 3 \\   
    \end{tabular}
    \end{footnotesize}
    \label{tab:example_results}
\end{table*}

\begin{figure*}[!ht]
    \centering
    \subfloat[\label{fig:invivo_bmode1_1540} Uncorrected Bmode.]{
        \includegraphics[trim ={8cm 4cm 4cm 4cm}, clip, width=.3\linewidth]{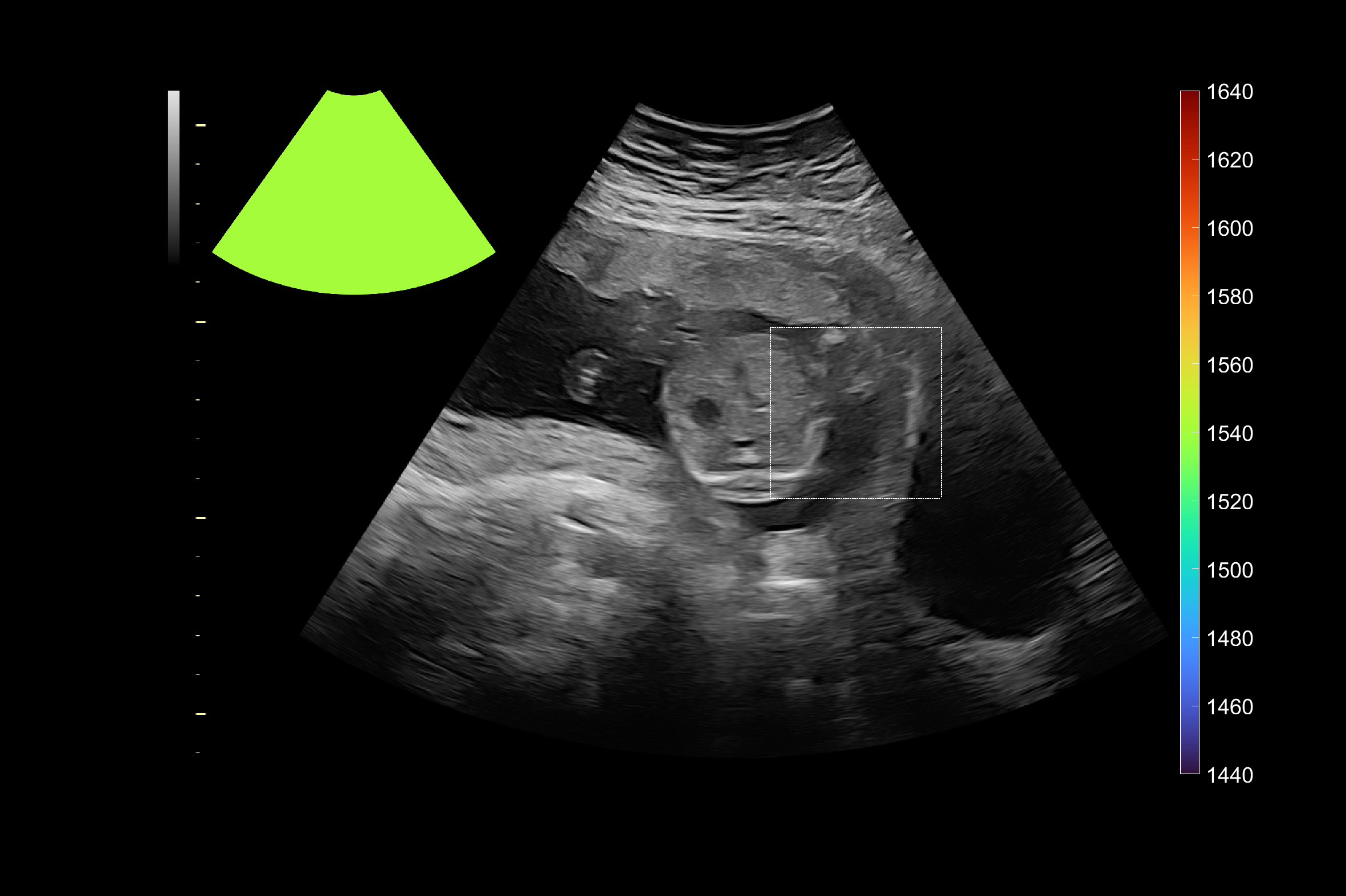}}
    \subfloat[\label{fig:invivo_bmode1_corrected} Aberration corrected Bmode.]{
        \includegraphics[trim ={8cm 4cm 4cm 4cm}, clip, width=.3\linewidth]{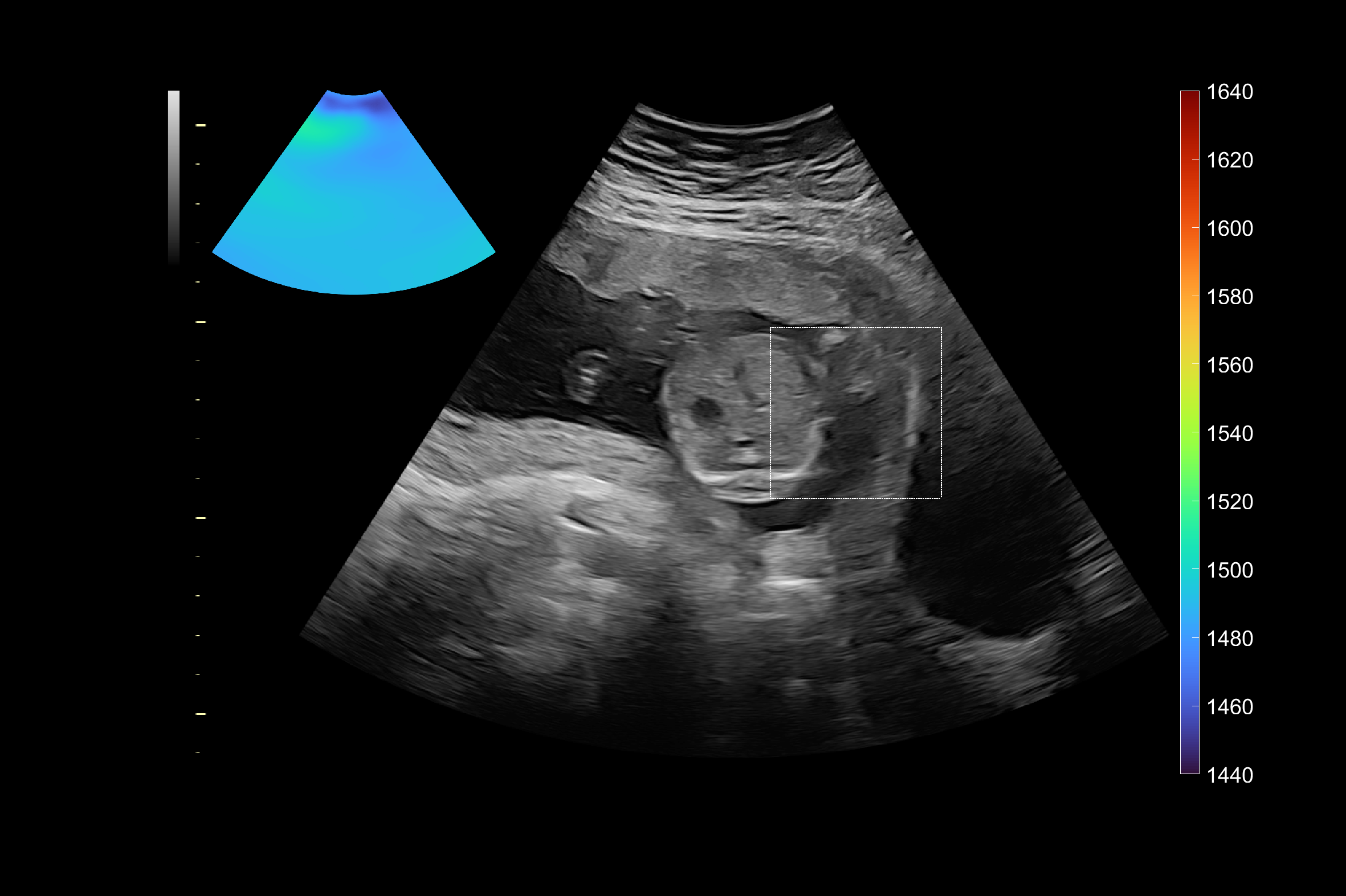}}
    \subfloat[\label{fig:invivo_bmode1_1540_zoom} Uncorrected Bmode.]{
        \includegraphics[trim ={0cm 0cm 21.5cm 0cm}, clip, width=.2\linewidth]{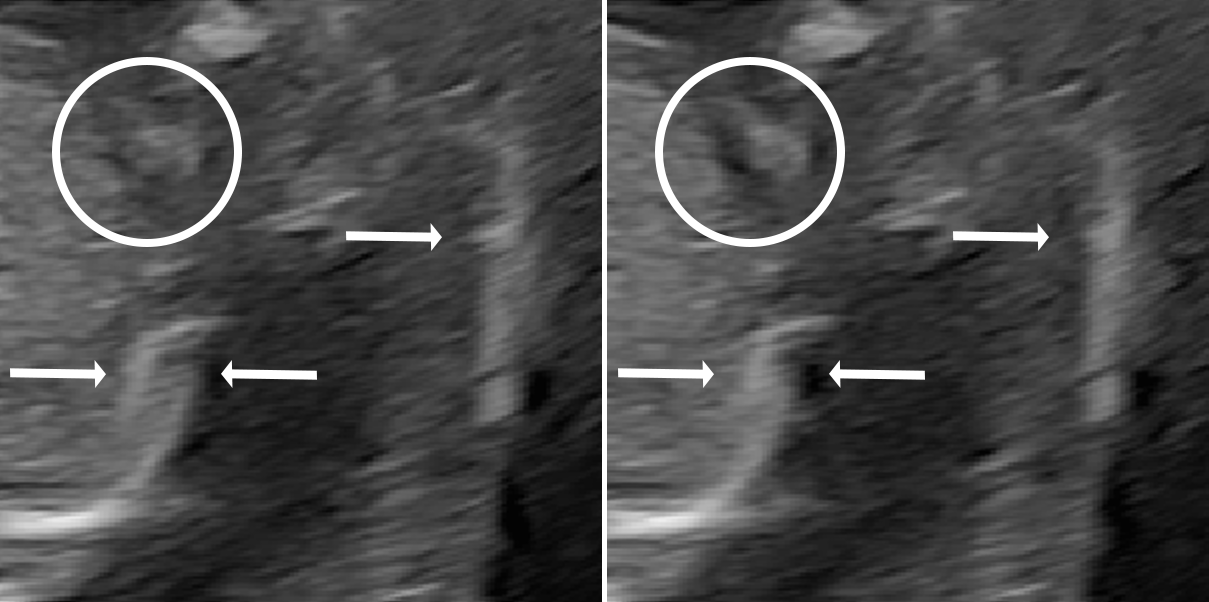}}
    \subfloat[\label{fig:invivo_bmode1_corrected_zoom} Corrected Bmode.]{
        \includegraphics[trim ={21.5cm 0cm 0cm 0cm}, clip, width=.2\linewidth]{Figures/Qualitative_results/with_arrows/20240614_101111_02S2_22_AFT_const_02S2_22_AFT_const.png}}
    \caption{Example A. \textit{In vivo} fetal Bmode image using constant \SI{1540}{\meter\per\second} and the estimated average sound speed map. The average sound speed map is shown in the top right corner and the rightmost colorbar indicates the sound speed values. Alternating GIFs are found in the supplementary material.}
    \label{fig:invivo1}
\end{figure*}

\begin{figure*}[!ht]
    \centering
    \subfloat[\label{fig:invivo_bmode2_1540} Uncorrected Bmode.]{
        \includegraphics[trim ={8cm 4cm 4cm 4cm}, clip, width=.3\linewidth]{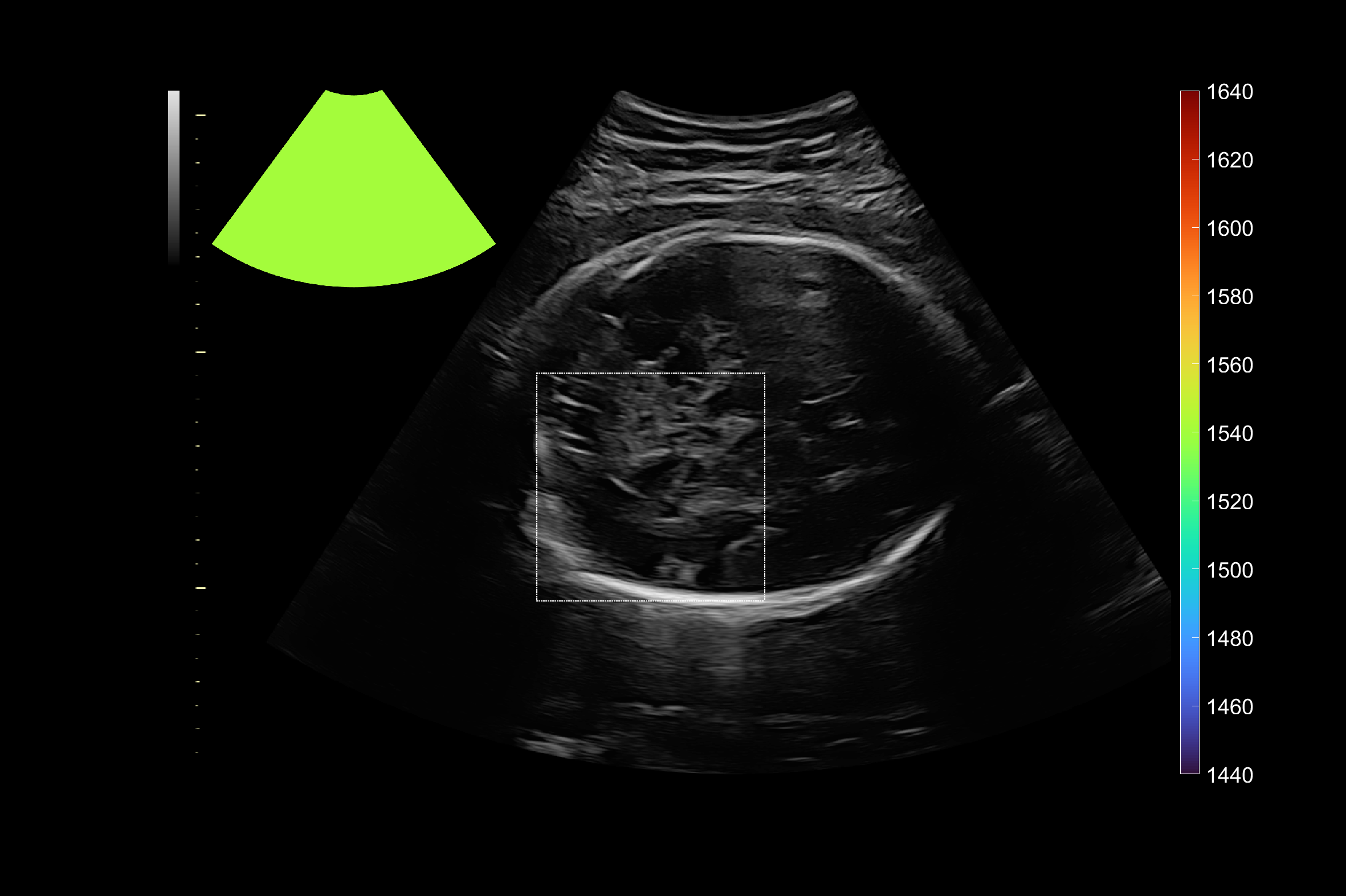}}
    \subfloat[\label{fig:invivo_bmode2_corrected} Aberration corrected Bmode.]{
        \includegraphics[trim ={8cm 4cm 4cm 4cm}, clip, width=.3\linewidth]{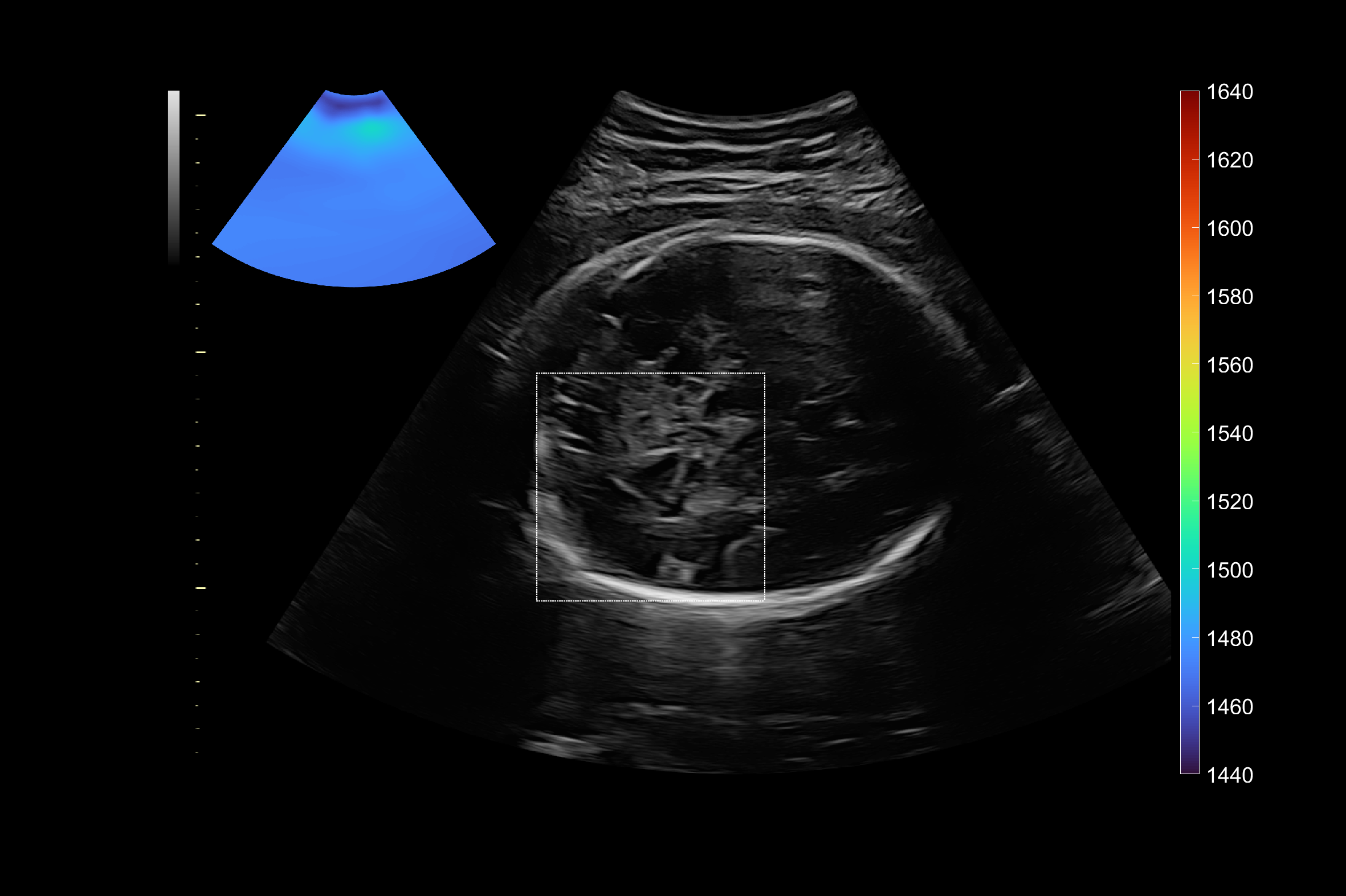}}
    \subfloat[\label{fig:invivo_bmode2_1540_zoom} Uncorrected Bmode.]{
        \includegraphics[trim ={0cm 0cm 28.5cm 0cm}, clip, width=.2\linewidth]{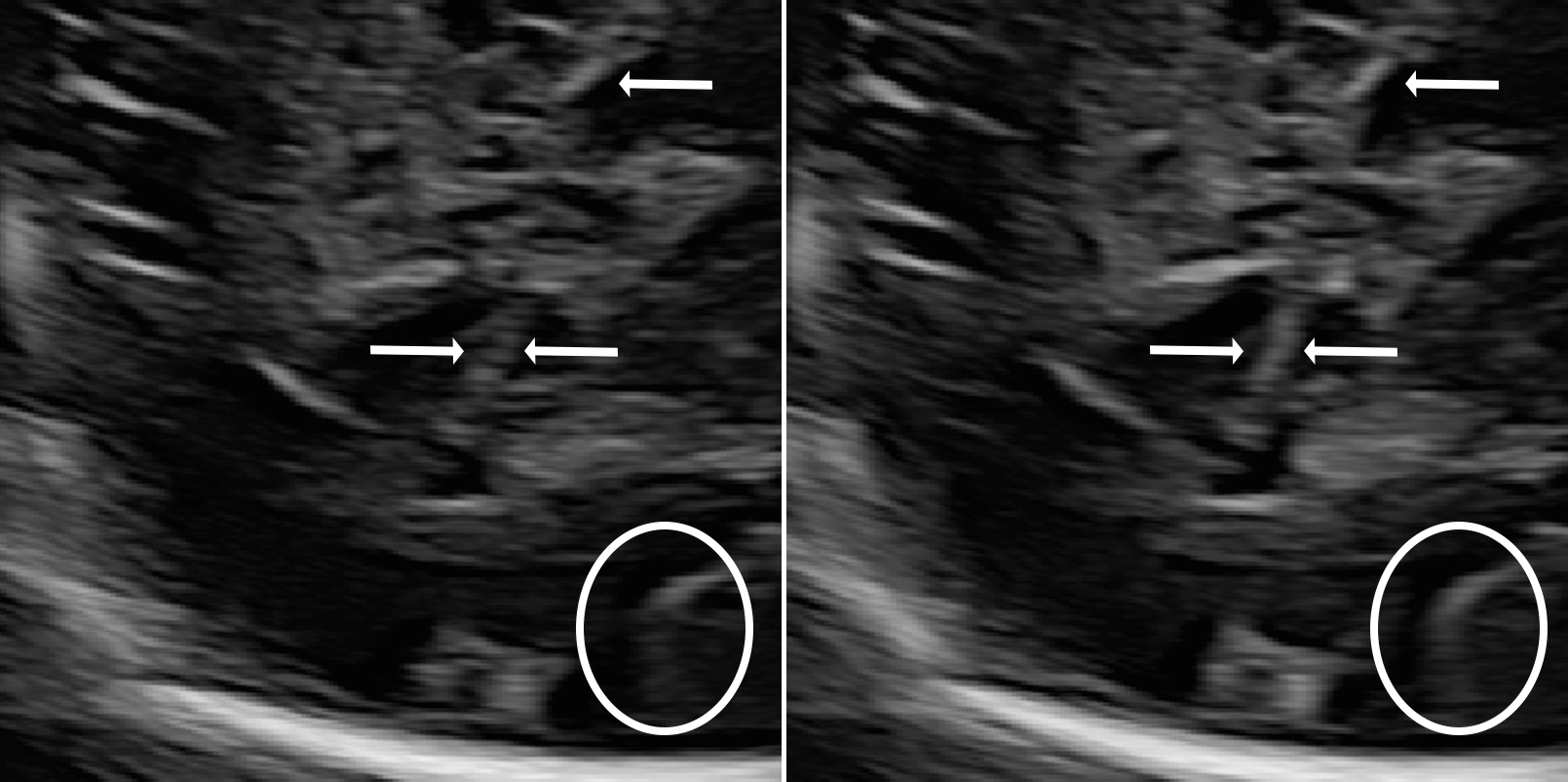}}
    \subfloat[\label{fig:invivo_bmode2_corrected_zoom} Corrected Bmode.]{
        \includegraphics[trim ={28.5cm 0cm 0cm 0cm}, clip, width=.2\linewidth]{Figures/Qualitative_results/with_arrows/20240614_100540_03S3_16_AFT_const_03S3_16_AFT_const.png}}
    \caption{Example B. \textit{In vivo} fetal Bmode image using constant \SI{1540}{\meter\per\second} and the estimated average sound speed map. The average sound speed map is shown in the top right corner and the rightmost colorbar indicates the sound speed values. Alternating GIFs are found in the supplementary material.}
    \label{fig:invivo2}
\end{figure*}

\begin{figure*}[!ht]
    \centering
    \subfloat[\label{fig:invivo_bmode3_1540} Uncorrected Bmode.]{
        \includegraphics[trim ={14cm 4cm 10cm 3cm}, clip, width=.28\linewidth]{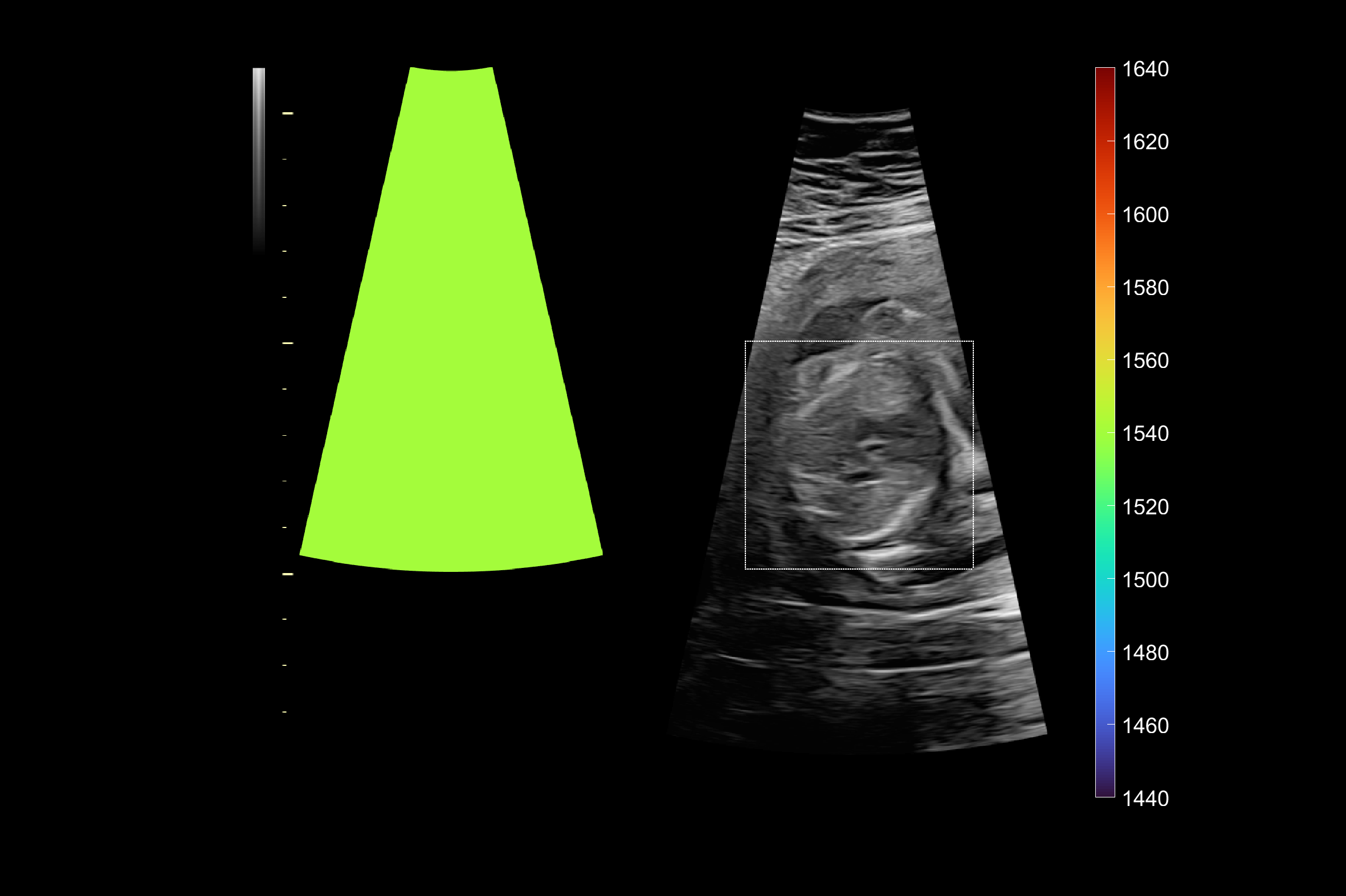}}
    \subfloat[\label{fig:invivo_bmode3_corrected} Aberration corrected Bmode.]{
        \includegraphics[trim ={14cm 4cm 10cm 3cm}, clip, width=.28\linewidth]{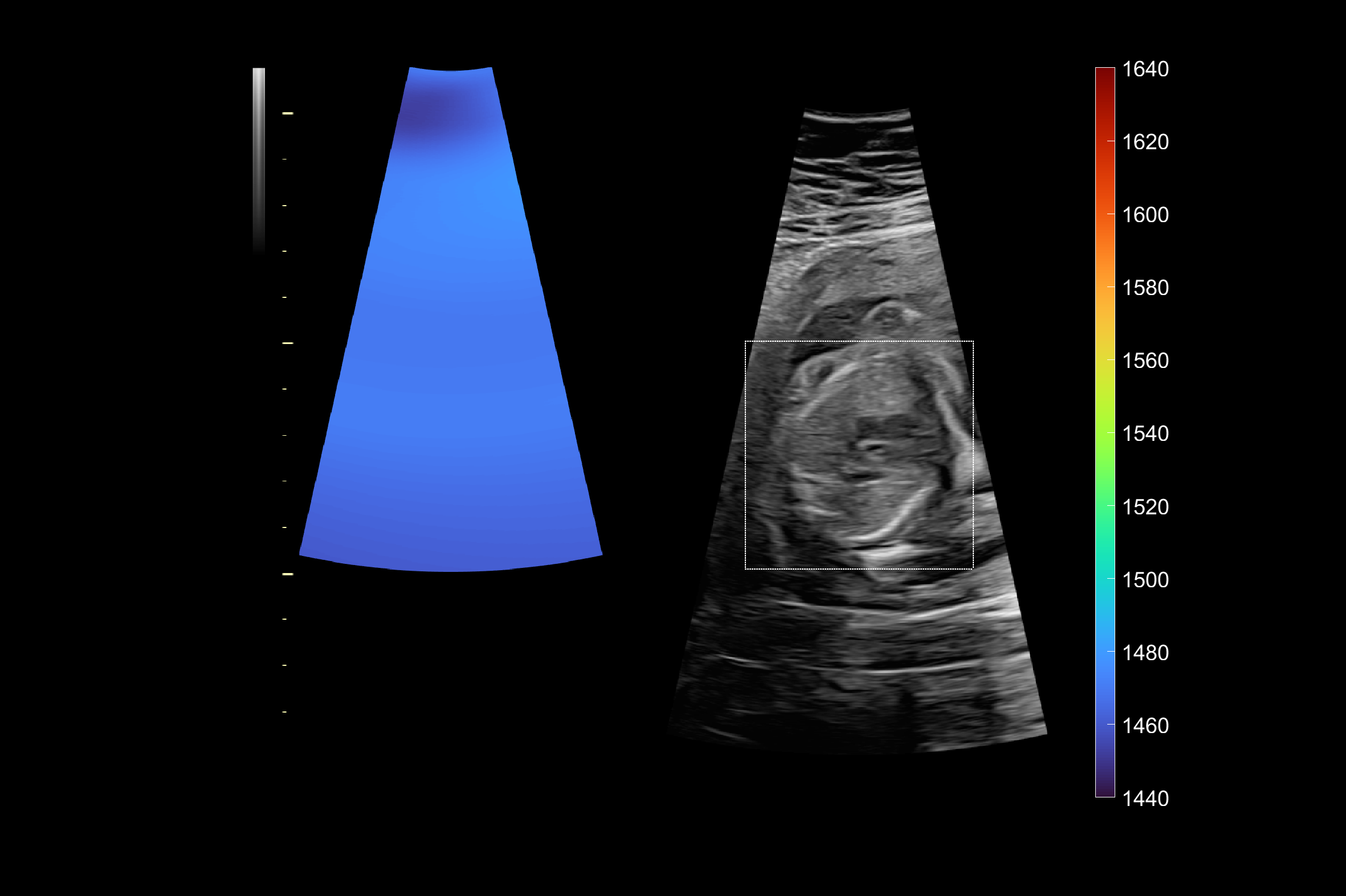}}
    \subfloat[\label{fig:invivo_bmode3_1540_zoom} Uncorrected Bmode.]{
        \includegraphics[trim ={0cm 0cm 28.5cm 0cm}, clip, width=.22\linewidth]{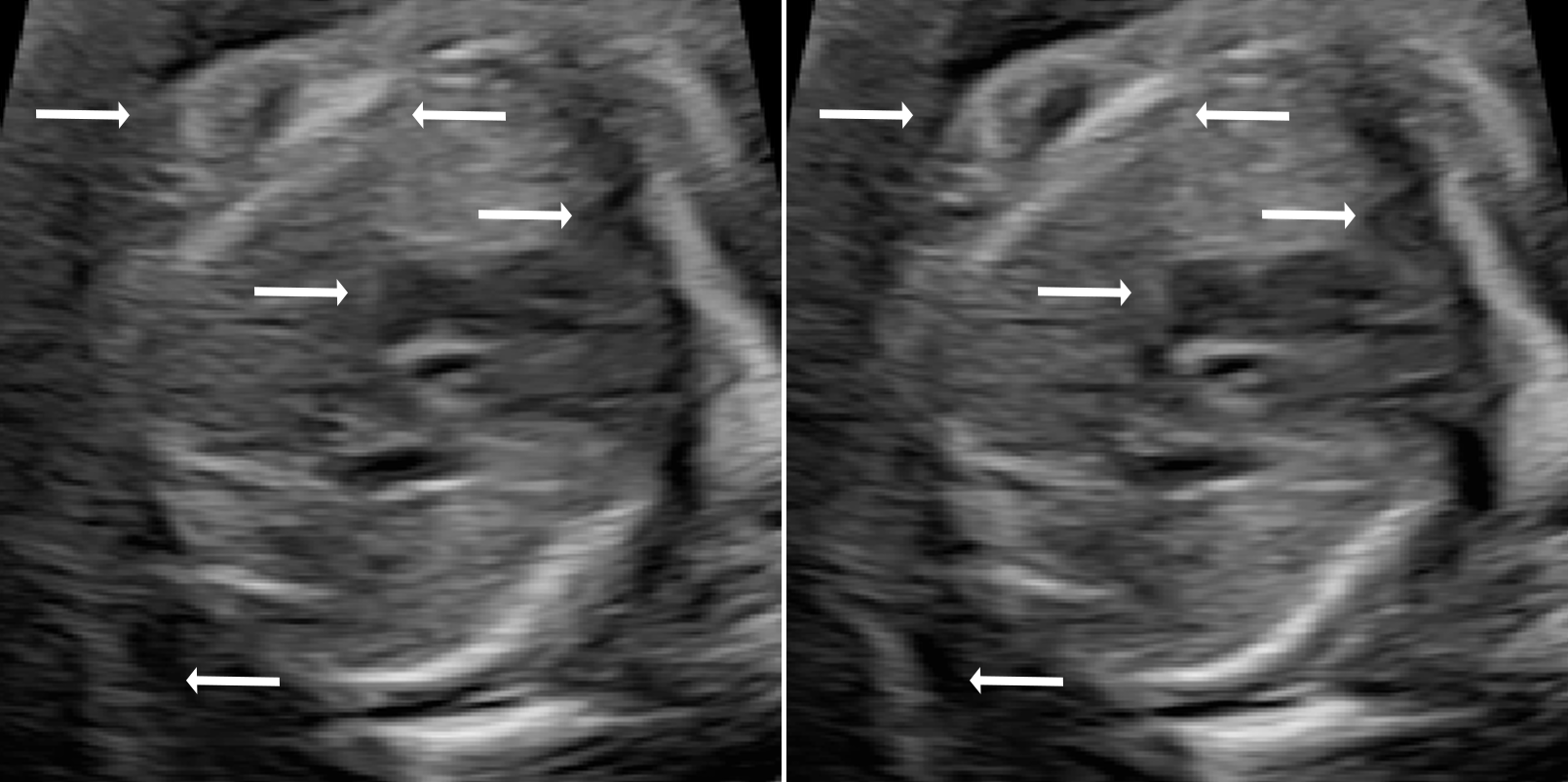}}
    \subfloat[\label{fig:invivo_bmode3_corrected_zoom} Corrected Bmode.]{
        \includegraphics[trim ={28.5cm 0cm 0cm 0cm}, clip, width=.22\linewidth]{Figures/Qualitative_results/with_arrows/20240614_102716_02S2_31_AFT_const_02S2_31_AFT_const.png}}
    \caption{Example C. \textit{In vivo} fetal Bmode and coherence images using constant \SI{1540}{\meter\per\second} and the estimated average sound speed map. The average sound speed map is shown in the top right corner and the rightmost colorbar indicates the sound speed values. Alternating GIFs are found in the supplementary material.}
    \label{fig:invivo3}
\end{figure*}

Three \textit{in vivo} Bmode fetal images with and without aberration correction are presented in Figs.~\ref{fig:invivo1}-\ref{fig:invivo3}. GIF images of examples A-L (see Fig.~\ref{fig:quanititative_sharpness} and Table~\ref{tab:example_results}) alternating between the corrected and uncorrected images are uploaded as supplementary material. The differences in image quality are easier to appreciate in the GIFs than in a side-by-side comparison. The data points for the three examples are colored black in Fig.~\ref{fig:quanititative_sharpness}. Table~\ref{tab:example_results} give tabular values for the global sound speed estimate, sharpness increase ($\kappa$) and answers from the clinical evaluations. 

Each aberration-corrected Bmode image is gained to have the same 80th percentile pixel intensity value as the 80th percentile value of the corresponding uncorrected image. The Bmode images are plotted in dB relative to the max of the uncorrected image. 

\section{Discussion}
\label{sec:Discussion}
The results indicate that the sound speed estimation \textit{in vitro} and \textit{in silico} performs well and that the proposed sound speed aberration correction increase image sharpness \textit{in vivo} and is favored by clinicians.

The estimated average sound speed maps for the simulated phantoms with known local maps show similarity to the true average map in the constant and layered cases. The results are similar to experiments without point targets published by Ali \textit{et al.} in \cite{ali_local_2022}. 

The sound speed estimation of the two CIRS phantom recordings appears to be similar, indicating that the estimator is independent of the assumed sound speed during transmission. This is important for \textit{in vivo} estimation because a potentially wrong sound speed is assumed when calculating aperture delays for a focused transmission. The phantom is manufactured to have a constant \SI{1540}{\meter\per\second}, and deviations in estimated sound speed can be a consequence of phantom wear or inaccuracies with the proposed method. 

The quantitative measurement of image sharpness in Fig.~\ref{fig:quanititative_sharpness} shows a general increase in sharpness when correcting for a deviation in sound speed. The measured Pearson Correlation Coefficient (PCC) of 0.67 signifies a clear trend.

The distribution of the estimated sound speed values in Fig.~\ref{fig:Single_sound_speed_histogram} shows about \SI{40}{\meter\per\second} lower mean value than the conventional sound speed of \SI{1540}{\meter\per\second}. A similar finding was observed in a clinical research paper by Chauveau \textit{et al.}, where images assuming \SI{1480}{\meter\per\second} were graded with higher quality than the images focused assuming \SI{1540}{\meter\per\second} \cite{chauveau_improving_2018}. Based on this, a lower default sound speed value could be considered for fetal ultrasound.

The distribution labeled as "similar quality" in Fig.~\ref{fig:boxplot_sound_speeds} shows a mean estimated sound speed value closer to \SI{1540}{\meter\per\second} than the other distributions. This is expected because images with an estimated sound speed close to \SI{1540}{\meter\per\second} will have a similar appearance to the uncorrected \SI{1540}{\meter\per\second} image. Example L, found in the supplementary material, has an estimated sound speed of 1534\unit{\meter\per\second} and is evaluated by all evaluators as "similar quality".

The \textit{in vivo} examples A-C, in Figs.~\ref{fig:invivo1}-\ref{fig:invivo3} respectively, show different nuances of how sound speed aberration correction improves focusing quality. A noticeable improvement in Fig.~\ref{fig:invivo1} is the improved contrast and clarity of borders. The sound speed correction seems to remove the double structure indicated by the arrows in Figs.~\ref{fig:invivo_bmode1_1540_zoom}-\ref{fig:invivo_bmode1_corrected_zoom}. The skull border and the brain structures in Fig.~\ref{fig:invivo2} are also better defined and clearer after aberration correction. Example C in Fig.~\ref{fig:invivo3} is the image with the highest increase in Tenengrad $\kappa$, seen from Fig.~\ref{fig:quanititative_sharpness} and Table~\ref{tab:example_results}. The improvement in image quality is evident around the boundary of the fetus body and the correction sharpens the structures indicated by the arrows. The improved contrast between the layers in the shallow tissue layers is also noticeable. 

Some of the measured sharpness values decrease after the correction, as seen in Fig.~\ref{fig:quanititative_sharpness}. Example D, found in the supplementary material, is one of them. The sound speed correction of Example D seems to make structures more continuous, which results in less high frequent lateral variations in the image and a lower Tenengrad value. This is a weakness of the Tenengrad sharpness measurement. The clinical evaluations answered differently for this image, as seen in Table~\ref{tab:example_results}.

Example J, found in the supplementary material, shows images of a fetal head for which the three evaluators preferred the \SI{1540}{\meter\per\second} image. The sound speed correction seems to sharpen the skull and tissue outside the head, but seems to defocus the structures on the left-hand side of center inside the head. A potential fix for this case is to use a smaller kernel size for smoothing the sound speed map, enabling more rapid spatial variations in average sound speed. Example G shows similar mixed corrections, and the corrected image was preferred by one clinician and the uncorrected preferred by two. 

It is interesting to note the difference in the classification done by the evaluators. It is apparent that the evaluators have different opinions of what makes two images of "similar quality", as one evaluator classified 77 images as such and another only 6. This difference raises an interesting research topic for future work. 

Further research will investigate what causes clinical evaluators to prefer the \SI{1540}{\meter\per\second} over the estimated sound speed in 27.5\% of the cases. One hypothesis for this is that the proposed method may choose sound speeds that focuses the images well in most part of the image, but not particularly at some clinical relevant structure, as seen in Example J and G. 2D matrix arrays can dynamically focus in the elevation direction, and it is expected that focusing with the true sound speed will obtain a narrower main lobe and lower sidelobes in elevation. This can explain some of the structure changes seen between corrected and uncorrected images and may be a reason for clinicians to prefer the uncorrected image. 

\section{Conclusion}
\label{sec:Conclusion}
This work shows that retrospective two-way aberration correction, based on coherence maximization of the distributed sound speed, is feasible and enhances image quality in fetal imaging. The increase in image quality is quantified for 172 images with a metric from digital photography, and the results show that sharpness increases when sound speed correction is performed. The aberration correction improves resolution and contrast, yielding clearer structures, better-defined borders, and sharper edges. The clinical evaluation shows that the proposed method is preferred or creates an image quality similar to \SI{1540}{\meter\per\second} images for 72.5\% of the 172 cases. The simulation and phantom studies show that the method is unbiased against the assumed sound speed on transmission and that the proposed aberration correction technique is equivalent to focusing using the ground truth distributed average sound speed map. For the data collected in this study, a sound speed of around \SI{1500}{\meter\per\second} is found to be a better choice for fetal imaging. Ultrasound systems capable to adapt to the anatomy of the individual patient, can enhance fetal imaging quality and contribute to better diagnostic precision.

\printbibliography
\section*{Acknowledgment}
The authors would like to thank Magnus Dalen Kvalevåg for support with the implementation in \textit{vbeam}. He also thanks Kjell Å. Salvesen for participating in the clinical evaluation. 

\end{document}